\documentclass{aastex63}

\usepackage{amsmath}

\begin{document}

\title{Effects of magnetic field orientations in dense cores on gas kinematics in protostellar envelopes}

\author[0000-0002-9959-1933]{Aashish Gupta}
\affiliation{Graduate Institute of Astronomy, 
National Central University, 300 Zhongda Road, Zhongli, Taoyuan 32001, Taiwan}
\affiliation{Academia Sinica Institute of Astronomy and Astrophysics, No. 1, Sec. 4, Roosevelt Road, Taipei 10617, Taiwan}
\affiliation{European Southern Observatory, Karl-Schwarzschild-Str. 2, 85748 Garching bei München, Germany}

\author{Hsi-Wei Yen}
\affiliation{Academia Sinica Institute of Astronomy and Astrophysics, No. 1, Sec. 4, Roosevelt Road, Taipei 10617, Taiwan}

\author{Patrick Koch} 
\affiliation{Academia Sinica Institute of Astronomy and Astrophysics, No. 1, Sec. 4, Roosevelt Road, Taipei 10617, Taiwan}

\author{Pierre Bastien}
\affiliation{Centre de recherche en astrophysique du Qu\'{e}bec \& d\'{e}partement de physique, Universit\'{e} de Montr\'{e}al, C.P. 6128 Succ. Centre-ville, Montr\'{e}al, QC, H3C 3J7, Canada}

\author{Tyler L. Bourke}
\affiliation{SKA Observatory, Jodrell Bank, Lower Withington, Macclesfield, Cheshire SK11 9FT, UK}
\affiliation{Center for Astrophysics | Harvard \& Smithsonian, 60 Garden Street, Cambridge, MA, USA}

\author{Eun Jung Chung}
\affiliation{Department of Astronomy and Space Science, Chungnam National University, 99 Daehak-ro, Yuseong-gu, Daejeon 34134, Republic of Korea}

\author{Tetsuo Hasegawa}
\affiliation{National Astronomical Observatory of Japan, National Institutes of Natural Sciences, Osawa, Mitaka, Tokyo 181-8588, Japan}

\author{Charles L. H. Hull}
\affiliation{National Astronomical Observatory of Japan, NAOJ Chile, Alonso de Córdova 3788, Office 61B, 7630422 Vitacura, Santiago, Chile}
\affiliation{Joint ALMA Observatory, Alonso de Córdova 3107, Vitacura, Santiago, Chile}

\author{Shu-ichiro Inutsuka}
\affiliation{Department of Physics, Nagoya University, Aichi 464-8602, Japan}

\author[0000-0003-2815-7774]{Jungmi Kwon}
\affiliation{Department of Astronomy, University of Tokyo, 7-3-1 Hongo, Bunkyo-ku, Tokyo 113-0033, Japan}

\author{Woojin Kwon}
\affiliation{Department of Earth Science Education, Seoul National University, 1 Gwanak-ro, Gwanak-gu, Seoul 08826, Republic of Korea}
\affiliation{SNU Astronomy Research Center, Seoul National University, 1 Gwanak-ro, Gwanak-gu, Seoul 08826, Republic of Korea}

\author{Shih-Ping Lai}
\affiliation{Academia Sinica Institute of Astronomy and Astrophysics, No. 1, Sec. 4, Roosevelt Road, Taipei 10617, Taiwan}
\affiliation{Institute of Astronomy and Department of Physics, National Tsing Hua University, Hsinchu 30013, Taiwan}

\author{Chang Won Lee}
\affiliation{Korea Astronomy and Space Science Institute, 776 Daedeokdae-ro, Yuseong-gu, Daejeon 34055, Republic of Korea}
\affiliation{University of Science and Technology, Korea (UST), 217 Gajeong-ro, Yuseong-gu, Daejeon 34113, Republic of Korea}

\author{Chin-Fei Lee}
\affiliation{Academia Sinica Institute of Astronomy and Astrophysics, No. 1, Sec. 4, Roosevelt Road, Taipei 10617, Taiwan}

\author{Kate Pattle}
\affiliation{Department of Physics and Astronomy, University College London, Gower Street, London WC1E 6BT, United Kingdom}

\author{Keping Qiu}
\affiliation{School of Astronomy and Space Science,  Nanjing University, 163 Xianlin Avenue, Nanjing 210023, P. R. China}
\affiliation{Key Laboratory of Modern Astronomy and Astrophysics (Nanjing University), Ministry of Education, Nanjing 210023, P. R. China}

\author[0000-0001-8749-1436]{Mehrnoosh Tahani}
\affiliation{Dominion Radio Astrophysical Observatory, Herzberg Astronomy and Astrophysics Research Centre, National Research Council Canada, P. O. Box 248, Penticton, BC V2A 6J9 Canada}

\author[0000-0002-6510-0681]{Motohide Tamura}
\affiliation{Department of Astronomy, University of Tokyo, 7-3-1 Hongo, Bunkyo-ku, Tokyo 113-0033, Japan}
\affiliation{Astrobiology Center, 2-21-1 Osawa, Mitaka-shi, Tokyo 181-8588, Japan}
\affiliation{National Astronomical Observatory, 2-21-1 Osawa, Mitaka-shi, Tokyo 181-8588, Japan}

\author{Derek Ward-Thompson}
\affiliation{Jeremiah Horrocks Institute, University of Central Lancashire, Preston PR1 2HE, UK}


\begin{abstract}
    
        Theoretically, misalignment between the magnetic field and rotational axis in a dense core is considered to be dynamically important in the star formation process, however, extent of this influence remains observationally unclear. For a sample of 32 Class 0 and I protostars in the Perseus Molecular Cloud, we analyzed gas motions using C$^{18}$O data from the SMA MASSES survey and the magnetic field structures using 850 $\micron$ polarimetric data from the JCMT BISTRO-1 survey and archive. We do not find any significant correlation between the velocity gradients in the C$^{18}$O emission in the protostellar envelopes at a 1,000 au scale and the misalignment between the outflows and magnetic field orientations in the dense cores at a 4,000 au scale, and there is also no correlation between the velocity gradients and the angular dispersions of the magnetic fields. However, a significant dependence on the misalignment angles emerges after we normalize the rotational motion by the infalling motion, where the ratios increase from $\lesssim1$ to $\gtrsim1$ with increasing misalignment angles. This suggests that the misalignment could prompt angular momentum transportation to the envelope scale but is not a dominant factor in determining the envelope rotation, and other parameters, like mass accretion in protostellar sources, also play an important role. These results remain valid after taking into account projection effects. The comparison between our estimated angular momentum in the protostellar envelopes and the sizes of the known protostellar disks suggests that significant angular momentum is likely lost between radii of $\sim$1,000-100 au in protostellar envelopes.

\end{abstract}

\section{Introduction}

Protostellar disks play a crucial role in both star and planet formation but details of how these disks form remain unclear. Historically, the disk formation was understood as a simple consequence of angular momentum conservation \citep{bod95}. However, the picture becomes more complicated when the magnetic field is considered. Ideal magnetohydrodynamics (MHD) simulations have shown that as a rotating magnetized dense core collapses, the infalling material drags the magnetic field inward, pinching the magnetic field lines and thereby, greatly increasing magnetic tension within the protostellar envelope. This allows the magnetic field to transport a significant amount of angular momentum outward, known as magnetic braking, and suppress the formation of a disk \citep{all03,gal06,mel08}. 


Several ideal MHD simulations of collapse of dense cores have suggested that inclusion of misalignment between the magnetic field and rotational axis of a dense core can greatly alter the final configuration of the magnetic field and reduce the efficiency of magnetic braking \citep{hen09,joo12,li13}. Some simulations have also suggested that the misalignment can instead increase the efficiency of magnetic braking \citep{mat04,tsu18}. However, \citet{hir20} demonstrated that this is likely because they simulate the very early accretion phase. Alternatively, non-ideal MHD simulations suggest that non-ideal MHD effects, namely ohmic dissipation, ambipolar diffusion, and the Hall effect, can greatly reduce the accumulation of magnetic flux in the inner region and thus, also allow the formation of a rotationally supported disk around a protostar \citep{inu10,mac14,mas16,wur19,hir20}. \citet{hir20} further show in their simulations with non-ideal MHD effects that the misalignment promotes the formation of larger disks in the later phase.

Observationally, the role of this misalignment on angular momentum transfer is not yet well understood.
For a sample of $\sim20$ protostars, \citet{gal20} compared the misalignments between the magnetic fields in the protostellar envelopes at a few thousand au scale and the outflow axes with the magnitudes of the velocity gradients in the protostellar envelopes at a $\sim5,000$ au scale, where the outflow axes were adopted as a proxy for the rotational axes of the protostellar sources. They found a positive correlation between the misalignment and the velocity gradient, which could suggest that a larger misalignment reduces the efficiency of magnetic braking. On the other hand, \citet{yen21arxiv} compared the sizes and fluxes of a sample of $\sim50$ protostellar disks observed in the 0.87 mm continuum emission with misalignment between their rotational axes and core-scale magnetic fields and found no significant correlations. This could suggest that misalignment does not play a crucial role in disk formation.

To investigate how dynamically important misalignment between the magnetic field and rotational axis in a dense core is in the star formation process, we studied $\sim1,000$ au envelope-scale kinematics in synergy with $\sim4,000$ au core-scale magnetic field orientations, for a sample of $\sim$32 Class 0 and I protostars in the Perseus cloud. The gas kinematics was analysed using C$^{18}$O (2--1) data at a resolution of $\sim$600~au taken by the Mass Assembly of Stellar Systems and their Evolution with the Submillimeter Array (SMA) survey (MASSES) \citep{ste19}, as described in Section \ref{sec:kinmatics}. The MASSES survey also measured outflow orientations in a subset of their sample \citep{ste17}, which can be taken as a proxy for the rotational axes of these systems \citep[e.g.,][]{cia10}. We compared the outflow orientations with the magnetic field orientations inferred from the $850~\micron$ polarization data at a resolution of $\sim3500$~au taken by the B-fields In STar-forming Region Observations (BISTRO) survey with the James Clerk Maxwell Telescope (JCMT) \citep{war17,cou19,doi20} as well as regular projects, as described in Section \ref{sec:field}. Comparisons between the gas kinematics and the magnetic field morphology, before and after accounting for projection and measurement uncertainties, are discussed in Section \ref{sec:InAnal} and \ref{sec:FiAnal}, respectively. Finally, Section \ref{sec:discussion} discusses possible physical interpretations of our key findings, and Section \ref{sec:conclusion} concludes this paper.

\section{Data}

\subsection{Sample Selection}
\label{sec:sample}

The sample of this study is selected from the SMA MASSES survey \citep{ste19}. This survey observed 1.3 mm and 850 $\micron$ continuum emission and several molecular lines towards 74 known Class 0 and I protostars in the Perseus molecular cloud. In a subset of 57 sources, the CO outflows were detected, and the outflow orientations were measured \citep{ste17}. The Perseus molecular cloud has also been observed with JCMT \citep{war17,cou19,doi20} to trace magnetic field structures with polarized submillimeter continuum emission. We selected sources with detections of outflows in CO and protostellar envelopes in C$^{18}$O with SMA and polarized 850 $\mu$m continuum emission within a radius of $4,000$~au with JCMT. These led to a sample of 32 sources.


\subsection{Gas Kinematics} \label{sec:kinmatics}

C$^{18}$O can trace protostellar envelopes \citep{oha97,gau20}. In this study we use C$^{18}$O ($2-1$) emission line data taken as part of the MASSES survey \citep{ste19} to analyse the gas kinematics in the protostellar envelopes at a $\sim1,000$~au scale. These data sets have a spectral resolution of $\sim 0.2$~km~s$^{-1}$, a spatial resolution of $\sim 2\arcsec$ ($\sim600~$au), and a maximum recoverable angular scale of $\sim 24\arcsec$ ($\sim7,000~$au). The details of the observations and the noise levels of the C$^{18}$O data are described in \citet{ste19}. The same data have also been used to study the morphology, flux, and velocity gradient of the C$^{18}$O emission in the protostellar envelopes at a larger scale by \citet{hem21}.

Using these data cubes, we constructed integrated intensity (moment 0) and intensity-weighted mean velocity (moment 1) maps as shown in Figure \ref{fig:maps1} and Figure \ref{fig:maps2}. To quantify the overall velocity gradients in the protostellar envelopes, we fitted the moment 1 maps with a two-dimensional linear model \citep{goo93}. We note that the velocity gradient in a protostellar envelope could change as a function of spatial scale. For a uniform comparison of the gas kinematics in our sample, all the velocity gradients were measured in the central regions within a radius of $1,000$~au in the sample sources, which is more than three times larger than the spatial resolutions.


The measured overall velocity gradients could trace a combination of rotational, infalling, and turbulent motions and even outflows \citep{gau20}. Assuming that a protostellar envelope is axisymmetric and the associated outflow is parallel to its rotational axis, the infalling and rotational motions tend to induce velocity gradients along and perpendicular to the outflow axis in the protostellar envelope, respectively \citep{yen13,pin19}.
Although a bipolar outflow also exhibits a velocity gradient along the outflow axis, the outflow is expected to have a different velocity structure from the infalling envelope, where the outflow and infalling velocities tend to decrease and increase with decreasing radii, respectively \citep{arc07}.
ALMA observations at higher resolutions of $\sim$100 au toward several protostars shows that the C$^{18}$O and CO emission lines trace different velocity structures at a 1,000 au scale, and the C$^{18}$O emission is more sensitive to infalling and rotating envelopes \citep{aso15,yen17,lee19}. 
Indeed we also found distinct velocity structures along the outflow axes in the C$^{18}$O ($2-1$) and CO ($3-2$) position-velocity (PV) diagrams using the SMA MASSES data (Figure \ref{fig:pv2}, Appendix \ref{sec:PVs}).
Thus, we extracted velocity profiles along and perpendicular to the outflow axes from the moment 1 maps and measured velocity gradients, to assess infalling and rotational motions in the protostellar envelopes.
Our measured velocity gradients along and perpendicular to the outflow axis agree with the velocity structures seen in the PV diagrams (Figure \ref{fig:pv}, Appendix \ref{sec:PVs}).



\subsection{Magnetic Field} \label{sec:field}

We used the JCMT polarimetric data at 850 $\mu$m to measure magnetic field structures in the dense cores in the Perseus molecular cloud.
The JCMT polarimetric data were taken using the polarimeter POL-2 with the large program BISTRO survey \citep[M16AL004 and M17BL011;][]{war17,cou19,doi20} and the regular projects (M17AP073 and M17BP058; PI: W.~Kwon). The angular resolution of JCMT at 850 $\mu$m is $\sim14\arcsec$, corresponding to $\sim3500$~au.
We obtained the catalog of the Stokes Q and U intensities of the polarization detections above $3\sigma$ in the Perseus molecular cloud from \citet{yen20}. The polarization data reduction was done following the procedures in \citet{pat17}. The pixel size of one detection is 12\arcsec. Polarization angles were calculated as $0.5 \times arctan(U/Q)$, 
where U and Q are respective Stokes parameters. These position angles were further rotated by $90^{\circ}$ to infer magnetic field orientations.

MHD simulations \citep{joo12,li13,hir20} studied angular momentum transportation during the collapse of dense cores with different initial magnetic field orientations. Unlike the envelope-scale magnetic field, which likely gets significantly deformed by the collapse \citep{gir06,mau18,kwo19}, the core-scale magnetic field structures are expected to remain relatively unaffected. Therefore, these larger-scale magnetic field orientations, traced with the JCMT observations, are suitable to compare with those in the simulations. 

\begin{figure}[htbp]
  \centering
  \includegraphics[scale=0.8]{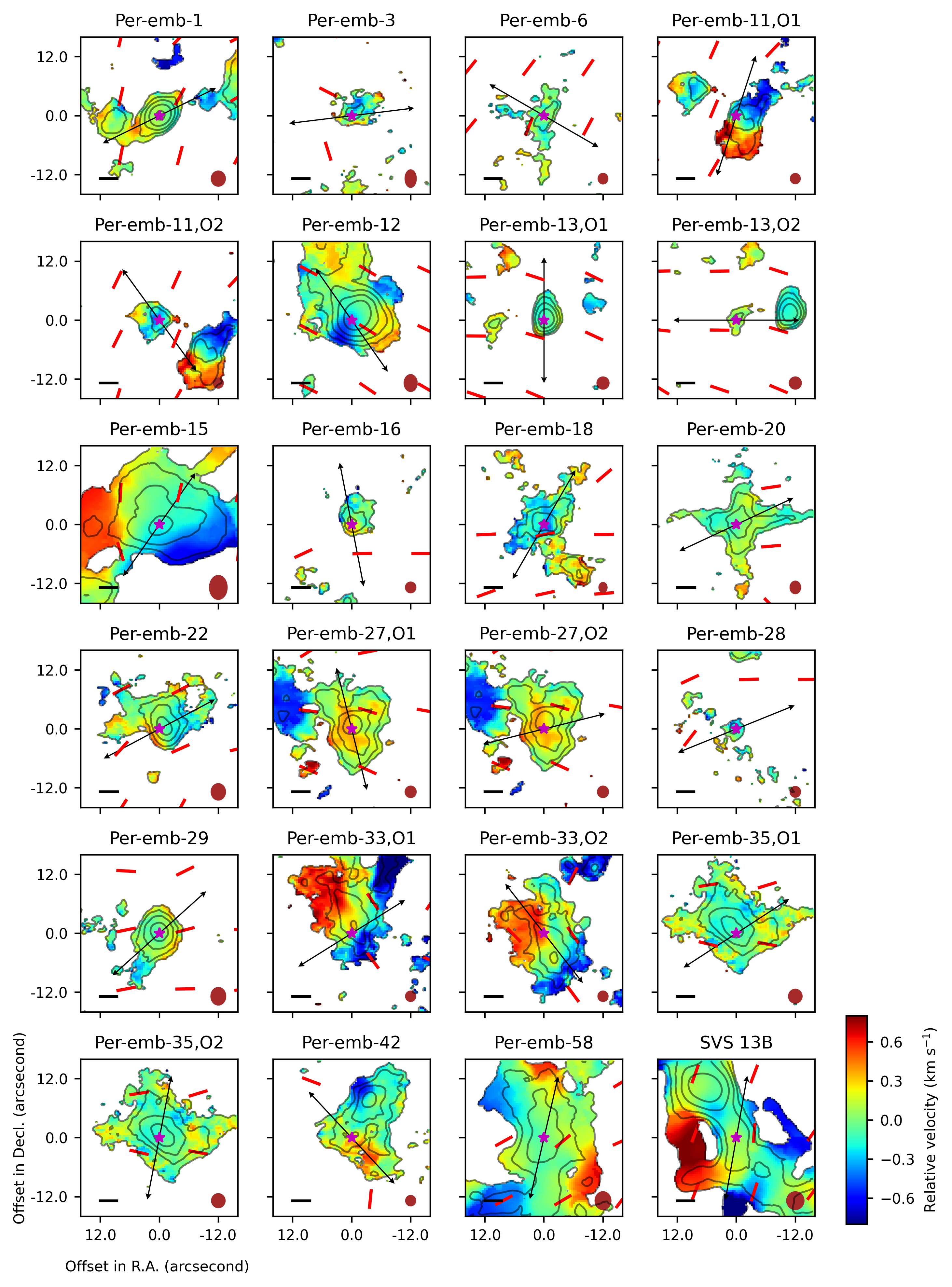}
    \caption{Magnetic field orientations observed with JCMT (red segments) overlaid on the C$^{18}$O moment 1 (color) and 0 (contours) maps obtained with the SMA observations. 
    The minimum separation between the red segments is 12$\arcsec$, comparable to the JCMT resolution of 14$\arcsec$ at 850~$\mu$m.
    Purple stars represent locations of the protostars and black arrows originating from them show outflow orientations.
    Black segments in the bottom-left corners depict a length scale of 1,000 au. Brown ellipses in the bottom-right corners depict beam sizes of the C$^{18}$O data.
    In each panel, the outermost contour represents the $3\sigma$ noise level in the moment 0 map and subsequent inner contours levels are increasing by a factor of two.}
  \label{fig:maps1}
\end{figure}

\begin{figure}[htbp]
  \centering
  \includegraphics[scale=0.8]{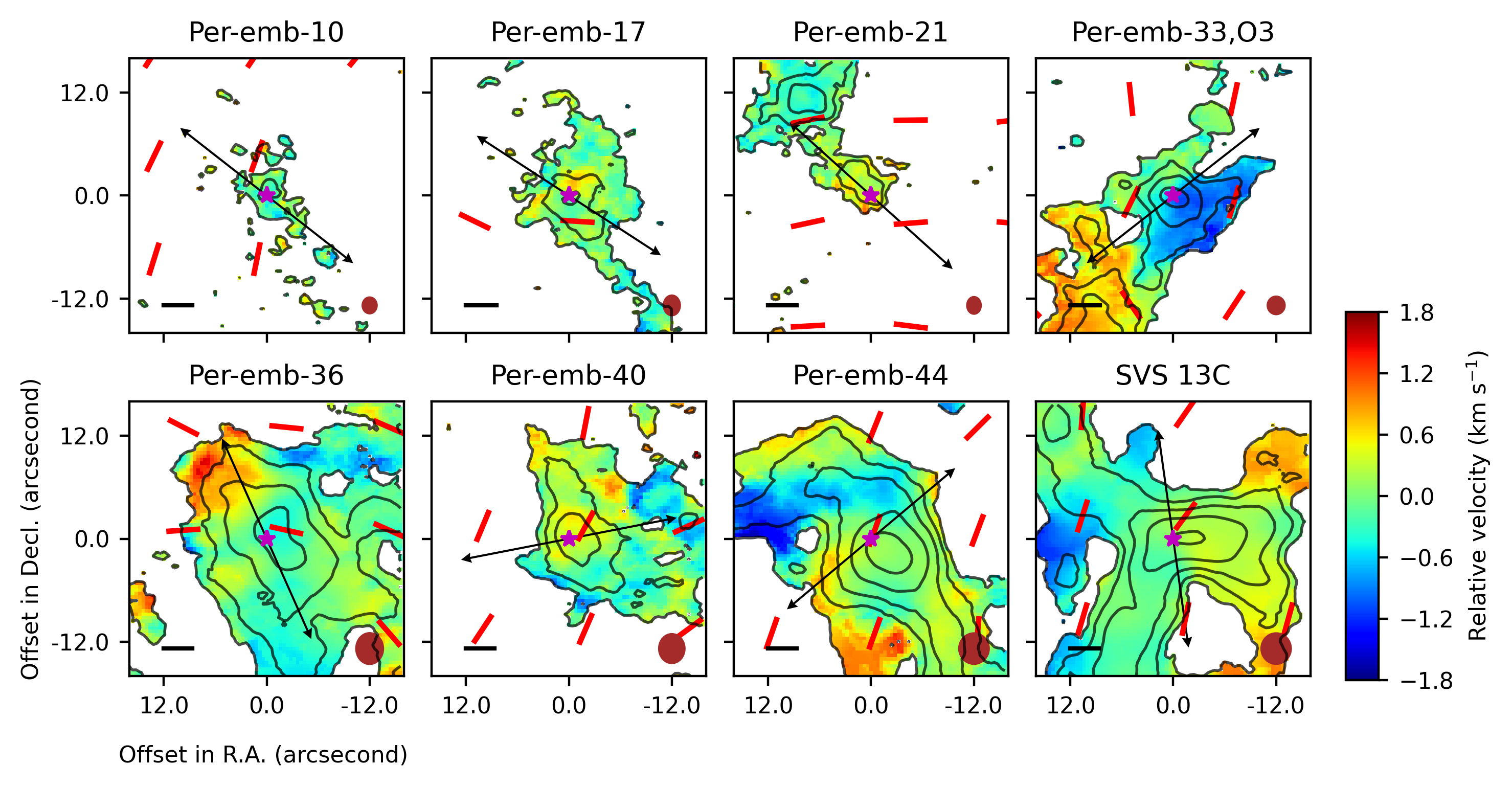}
    \caption{Same as Fig.~\ref{fig:maps1} but with a different color scale for the moment 1 maps.}
  \label{fig:maps2}
\end{figure}

Using the magnetic field information (as shown in Figure \ref{fig:maps1} and Figure \ref{fig:maps2}), we measured core-scale field orientations and angular dispersions. \citet{cur10} analysed dense cores in the Perseus region using 850 $\mu$m SCUBA maps and found the typical core radius to be $\sim0.02$~pc or $\sim4,000$~au. 
Thus, for each source in our sample, we first computed mean Stokes Q and U intensity from the detections within a radius of 4,000 au from the protostar and then estimated the overall orientation of the magnetic field. Standard deviation of the position angles of the individual field orientations was taken as the angular dispersion. According to the Davis–Chandrasekhar–Fermi method, this dispersion is proportional to the turbulence over the plane-of-sky magnetic field strength \citep{dav51,cha53}, and thus, can be considered as a proxy for the field strength which is an another important parameter in the MHD simulations. Errors in both of these quantities were estimated with error propagation. The typical error is $\lesssim 5\arcdeg$. 


All the measurements of the velocity gradients in the protostellar envelopes and the orientations and angular dispersions of the magnetic fields in the dense cores analyzed in the present paper are presented in Appendix \ref{sec:measurements}.


\section{Analysis and Results}

\subsection{Observed velocity gradients vs magnetic field structures}  
\label{sec:InAnal}

We inspected relations between the velocity gradients at the envelope scale (Section \ref{sec:kinmatics}) and the misalignment between the core-scale magnetic field orientations (Section \ref{sec:field}) and the outflows. We assume the rotational axes of the protostellar envelopes to be same as the outflow axes, and the MHD simulations suggest that they are indeed mostly parallel \citep{cia10,mac20}. We also note that the rotational axis of the central protostar-disk system, where outflows are launched, in a protostellar source might not be perfectly aligned with the rotational axis of its protostellar envelope. Observations have indeed found misaligned disks and protostellar envelopes around Class 0 and I protostars \citep{lee19,sai20}, which could suggest misaligned rotational axes of the disk and the envelope. Nevertheless, the observed misalignment angles between the disks and the protostellar envelopes are typically small and less than 10\arcdeg--20\arcdeg, which is comparable to the uncertainty in the outflow directions \citep{ste17}.

Three different velocity gradients in the protostellar envelopes are discussed in the present paper, namely (1) {\it overall velocity gradient}, which traces overall envelope-scale kinematics (hereafter overall gradient), (2) {\it velocity gradient perpendicular to the outflow}, which is expected to be proportional to rotational motion (hereafter rotational gradient), and (3) {\it velocity gradient parallel to the outflow}, which is expected to be proportional to infalling motion. In addition, we also computed normalized rotational gradients by dividing the rotational gradients by the velocity gradients parallel to the outflow axes, which are taken as a proxy for strength of the rotational motion for the given infalling motion in a protostellar envelope. We note that the velocity gradients perpendicular to and along the outflow axis might not completely trace rotational and infalling motions in a protostellar envelope \citep[e.g.,][]{tob12_grad}. Nevertheless, these velocity gradients can still be considered as upper limits of rotational and infalling velocities because faster rotational and infalling motions are expected to induce larger velocity gradients perpendicular to and along the outflow axis in a protostellar envelope \citep{yen13,pin19,gau20}. 

\begin{figure}[htbp]
  \centering
  \includegraphics[scale=0.8]{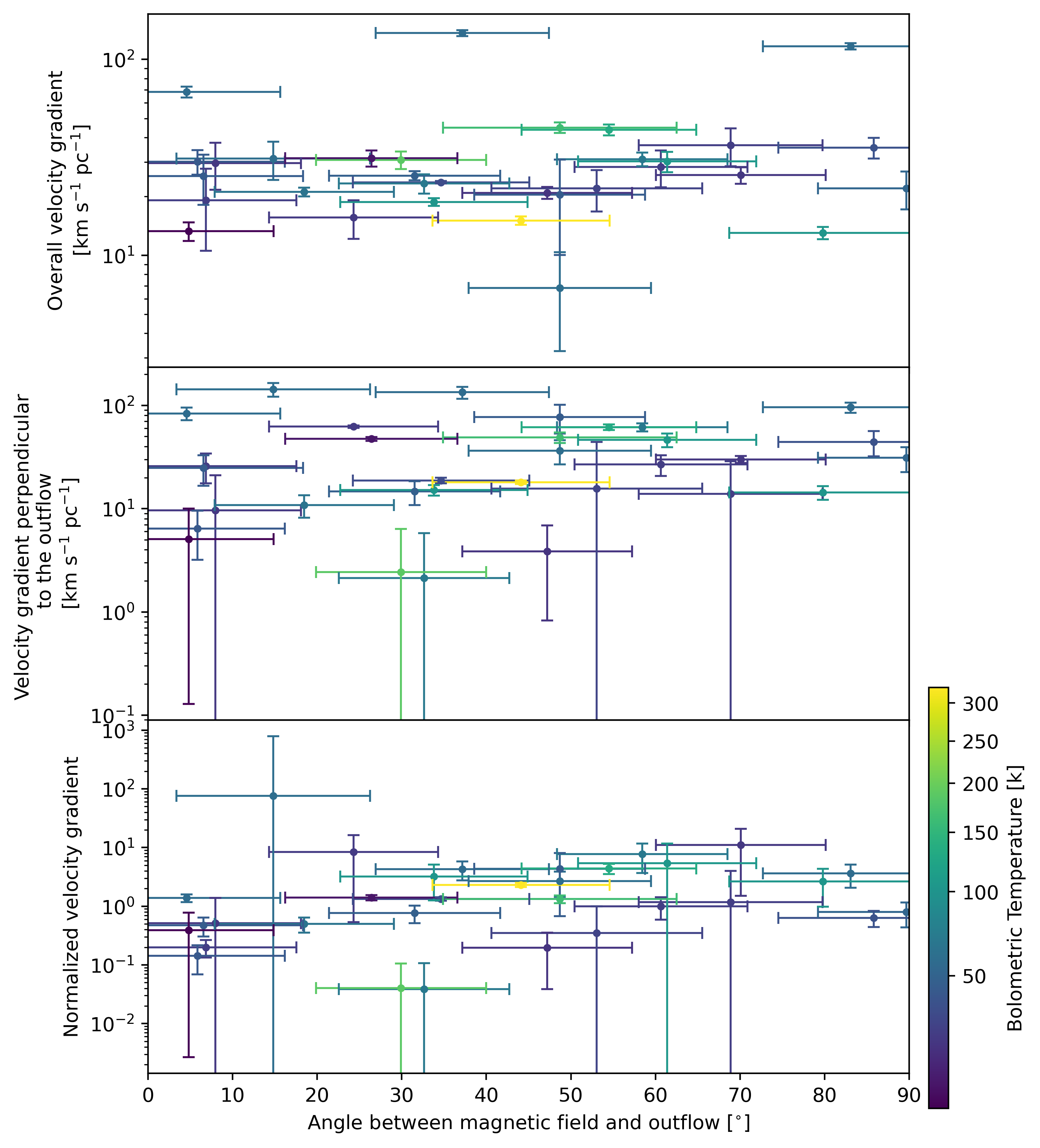}
    \caption{Misalignment between the magnetic field and outflow axis in dense cores as a function of overall velocity gradient (top panel), velocity gradient perpendicular to the outflow (middle panel), and velocity gradient perpendicular to the outflow divided by that along the outflow axis (bottom panel) in the protostellar envelopes at a 1,000 au scale. Marker colours represent bolometric temperatures of corresponding protostars.}
  \label{fig:corr1}
\end{figure}

Figure \ref{fig:corr1} shows the overall, rotational, and normalized rotational gradients in the protostellar envelopes as a function of the misalignment.
We use Spearman rank correlation analysis to study the correlation between the velocity gradients and misalignment angles. The Spearman correlation coefficients and corresponding p-values, as computed using \textit{scipy} package in Python are 0.12 and 0.52 for the overall gradient, 0.22 and 0.23 for the rotational gradient, and 0.35 and 0.05 for the normalized rotational gradient, respectively. 
P-values here refer to the probability of getting these correlation coefficients from a random uncorrelated sample, and a low p-value of $\lesssim 0.05$ suggests that the observed correlation is significant.
The coefficients and p-values for the overall and rotational gradients suggest that the gas kinematics in the protostellar envelopes at a 1,000 au scale does not strongly depend on the misalignment. In contrast, with a p-value of $0.05$, the normalized rotational gradient likely has a significant dependence on the magnetic field orientation. This can also be seen in Figure \ref{fig:corr1}, where unlike the overall (top panel) and rotational gradients (middle panel), the normalized rotational gradient (bottom panel) shows an increase from $\sim0.1$--$1$ for the projected magnetic fields roughly parallel to the outflow axes to $\sim1$--$10$ for nearly orthogonal configurations.

It is worth emphasizing that this correlation -- emerging only once the ratio is formed between the perpendicular velocity gradient and the parallel velocity gradient -- is transitioning from smaller-than one to larger-than one. As such, Figure \ref{fig:corr1}c is revealing two different regimes: a more infall-dominated regime with the magnetic field closely aligned with the outflow axis within $\sim 30^{\circ}$, and a more rotation-dominated regime where the ratios grow to larger than one, likely enabled by the larger misalignment angles. These different regimes are unnoticed in Figure \ref{fig:corr1}b when only the absolute magnitude of the rotational velocity gradient is considered. This demonstrates that for such studies it is crucial to form physically motivated quantities that can capture the dynamics in sources that differ in mass and velocity gradient. Possible physical interpretations of these observed trends are discussed in more detail in Section \ref{sec:MisVSVel}.

\begin{figure}[htbp]
  \centering
  \includegraphics[scale=0.8]{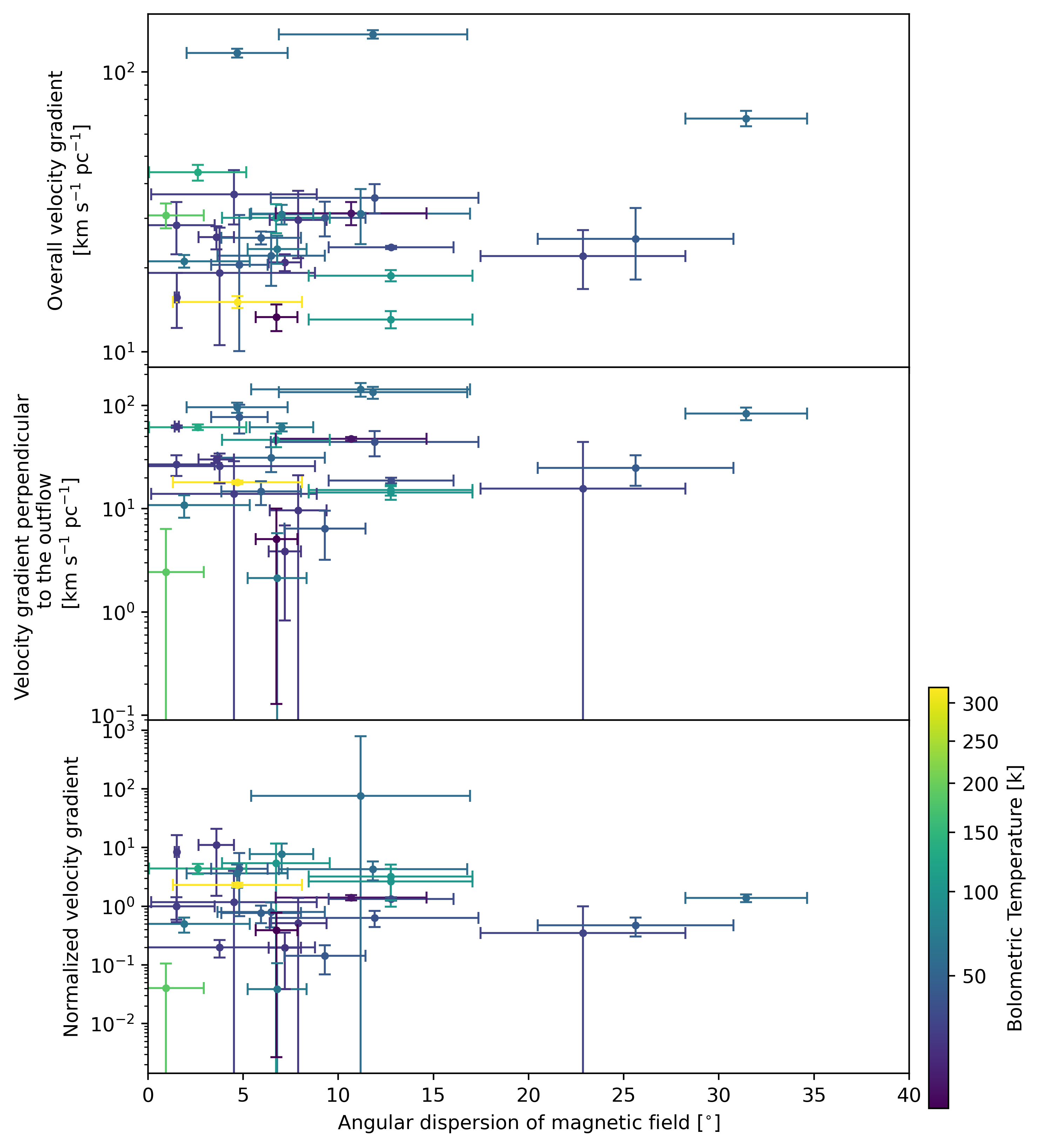}
    \caption{Angular dispersion of the magnetic field as a function of overall velocity gradient (top panel), velocity gradient perpendicular to the outflow (middle panel), and velocity gradient perpendicular to the outflow divided by that along the outflow axis (bottom panel) in the protostellar envelopes at a 1,000 au scale. Marker colours represent bolometric temperatures of corresponding protostars.}
  \label{fig:corr2}
\end{figure}

We also analysed these velocity gradients at a $1,000$~au scale as functions of the angular dispersion of the magnetic fields within $4,000$~au of the protostars, as shown in Figure \ref{fig:corr2}. The Spearman correlation coefficients and p-values for these relations are 0.06 and 0.74 for the overall gradient, 0.06 and 0.75 for the rotational gradient, and -0.07 and 0.72 for the normalized rotational gradient, respectively. Since all the p-values are $>0.7$, we did not find any significant dependence of the gas kinematics in the protostellar envelopes on the angular dispersion of the magnetic fields. In addition, we also found that there is no clear dependence of the velocity gradients and the orientations and angular dispersions of the magnetic fields on the bolometric temperatures of the protostars, which can be an evolutionary indicator for protostellar sources \citep{che95}.

\subsection{Corrections for projection effects}
\label{sec:corrs}

The simple correlation coefficient analysis in Section \ref{sec:InAnal}, does not account for uncertainties in the measurements of both the velocity gradients and the angles between the outflows and the magnetic fields. 
Besides the measurement uncertainties, the angles measured between the outflows and the magnetic field orientations are angles projected on the plane of the sky (POS), and the actual misalignment in three-dimensional (3D) space might differ significantly. \citet{gal20} demonstrated that the misalignment is likely underestimated due to the projection effect, and this effect is especially prominent for smaller misalignment angles ($\lesssim40^{\circ}$). As derived in Appendix \ref{sec:angles}, for a given projected angle, the actual angle in 3D space can be determined if the inclinations of the magnetic field and the outflow relative to POS are known, as
\begin{equation}
\label{equ:angle}
    \cos \theta = \cos \alpha \times \cos \beta \times \cos\lambda + \sin\alpha\times \sin\beta,
\end{equation}
where $\theta$ denotes the actual angle in 3D, $\lambda$ is the angle projected on POS, $\alpha$ denotes the inclination of the outflow with respect to POS, and $\beta$ denotes the inclination of the magnetic field with respect to POS. Because $\alpha$ and $\beta$, i.e. the 3D orientations of the magnetic fields and the outflows in our sample, are not known, we assume a probability distribution of these angles and estimate the underlying probability distribution of the actual angle ($\theta$) from the observed angle ($\lambda$).

The outflow axis and the magnetic field are more likely to be closer to POS in our sample. This is because (1) for a uniform distribution of unit vectors in 3D space more vectors will lie around the equator (parallel to POS) than near the pole (along the line of sight) and (2) we use observational results of magnetic field and outflow components projected on POS inheriting a bias against vectors perpendicular to POS. For simplicity, we generated the distributions of $\alpha$ and $\beta$ assuming that the outflow and magnetic field orientations are uniformly distributed in 3D space. For this, $\alpha$ and $\beta$ should follow a cosine distribution, i.e., $P(\beta) \propto \cos{\beta}$, where is $P(\beta)$ is the probability of the inclination of the magnetic field being equal to $\beta$. Nevertheless, the actual distributions of these angles are not known and can be different from our assumptions. We also repeated our analysis with differently assumed distributions and the final results do not change significantly, as quantified in Appendix \ref{sec:alphabeta}.

Similarly, we corrected the projection effects on the rotational gradients with the assumption of the probability distributions of $\alpha$. When a rotational axis is inclined with respect to POS ($\alpha>0^{\circ}$), the rotational gradient tends to be underestimated with observations. The difference between the actual and observed gradients increases when the protostellar envelope of a source is more face on. For a given $\alpha$, the actual rotational gradient (${\rm VG}_{true}$) can be estimated as,
\begin{equation}
\label{equ:rot}
    \text{VG}_{true} = \text{VG}_{obs}/\cos{\alpha}
\end{equation}
where $\text{VG}_{obs}$ is the rotational gradient observed in the moment 1 maps. We did not apply this correction for the normalized rotational gradient because the projection effects ($\cos{\alpha}$) on the velocity gradients parallel and perpendicular to the outflow axis cancels out.

The Perseus region has also been observed in the continuum emission at 8 mm, 1 cm, 4 cm, and 6.6 cm with the VLA Nascent Disk and Multiplicity (VANDAM) survey with the Karl G.\ Jansky Very Large Array (VLA), with angular resolutions down to $\sim 0.06\arcsec$ \citep{tob15}. For ten sources in our sample, their disks were resolved with the VANDAM survey, and their inclination angles were measured \citep{seg18}. For these sources, instead of using the assumed distributions, the disk inclination angles were converted to $\alpha$ values and used to constrain the probability distributions of $\theta$ and deprojected rotational gradients. We note that the disks and the envelopes could be misaligned. Nevertheless, adopting the disk inclinations would still provide better constraints than our simply assumed distribution because of the typically small misalignment angle (if present) of $<$10\arcdeg--20$\arcdeg$ between disks and envelopes \citep[e.g.,][]{lee19,sai20}.



\subsection{Deprojected velocity gradients and misalignment angles}
\label{sec:FiAnal}
In order to account for the measurement and systematic uncertainties discussed in Section \ref{sec:corrs}, we simulated expected probability distributions of deprojected velocity gradients and misalignment angles from our observational measurements. Firstly, for each measurement, we generated 10,000 simulated data points following a normal distribution with the observed values as means and their measurement uncertainties as standard deviations. Then we corrected the projection effects on these simulated data points. 
The probability distribution of deprojected misalignment angles ($\theta$) is estimated from the observed angles ($\lambda$) with Equation \ref{equ:angle} on the assumption of the probability distributions of $\alpha$ and $\beta$, as discussed in Section \ref{sec:corrs}. Similarly, the distributions of deprojected rotational gradients are also estimated from the distributions of the observed rotational gradients following Equation \ref{equ:rot}.

\begin{figure}[htbp]
\centering
  \includegraphics[scale=0.8]{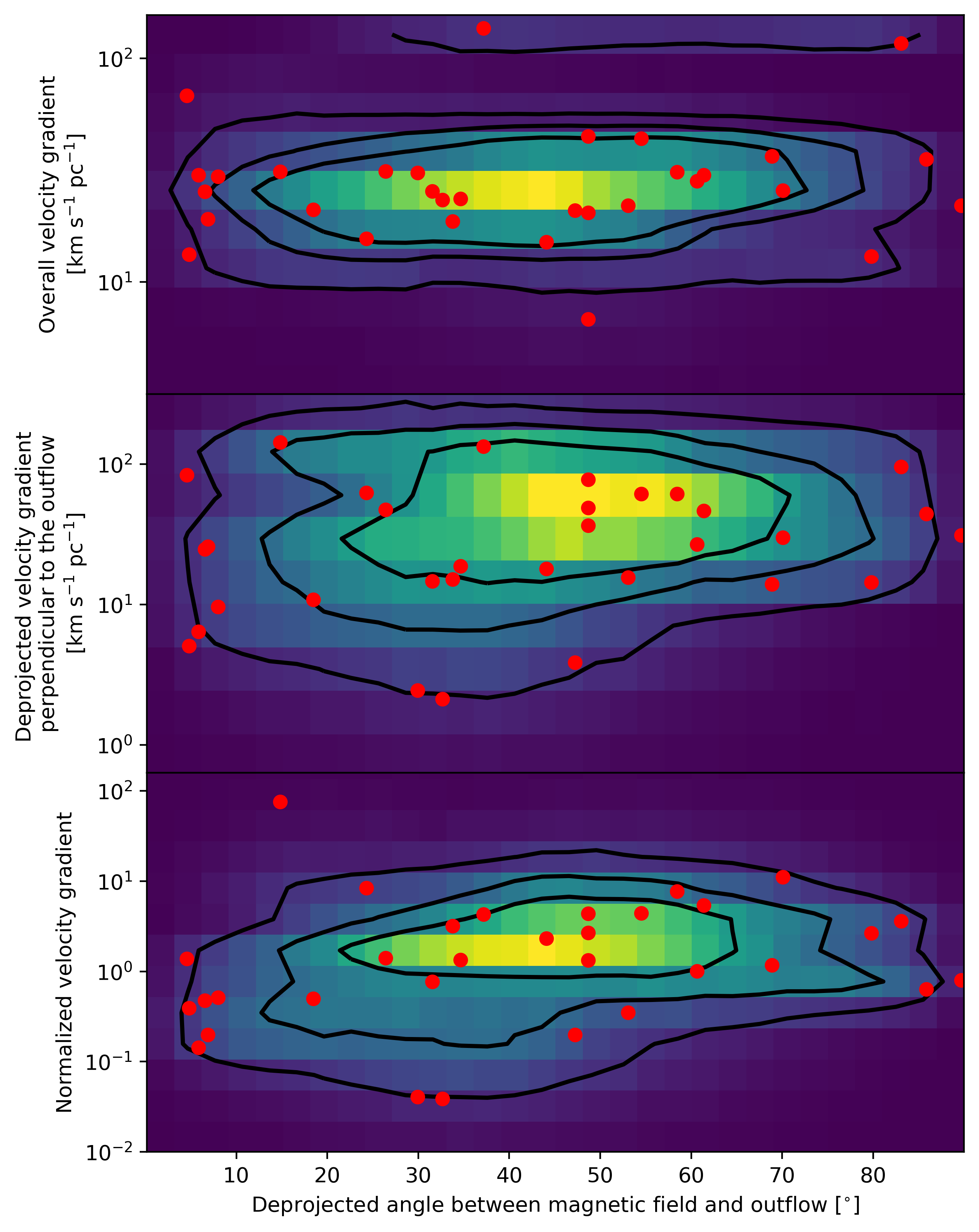}
    \caption{Simulated probability distributions of overall gradient (top panel), deprojected rotational gradient (middle panel), and normalized rotational gradient (bottom panel) in the protostellar envelopes with respect to deprojected misalignment angles between the core-scale magnetic fields and outflow axes, as discussed in Section \ref{sec:FiAnal}. Red circles represent the original measurements. Contours represent probability levels: 0.91, 0.68, 0.52 for the top panel, 0.89, 0.61, 0.38 for the middle panel, and 0.86, 0.60, 0.27 for the bottom panel. 
    }
  \label{fig:corr_deproj}
\end{figure}

Finally, we have 10,000 simulated values of the overall velocity gradient, deprojected rotational gradient, normalized rotational gradient, and deprojected misalignment for each of the 32 sources. Figure \ref{fig:corr_deproj} shows the inferred (or simulated) probability distributions of the velocity gradients with respect to the angles between the magnetic fields and the rotational axes in 3D space. The overall gradients (top panel) do not show any clear trend with respect to the misalignment. The rotational gradients (middle panel) seem to narrow down slightly to higher values with increase in the misalignment, as there are almost no small rotational gradients ($<10$~km~s$^{-1}$~pc$^{-1}$) when misalignment angles are $>50\arcdeg$. However, any overall trend is still not obvious, and the lack of data points in the bottom right corner could just be due to our limited sample size. The normalized rotational gradient (bottom panel) seems to display a more prominent positive trend with respect to the misalignment, with the typical ratio smoothly increasing from $<1$ for smaller misalignment angles to $>1$ for larger angles.


Quantitatively, the correlation coefficient analysis is not straight-forward for these simulated distributions as the number of data points is artificially too large and thus, p-values estimated using the standard methods will come out to be too small. Instead, we made groups of 32 simulated data points, where for each group we randomly pick one simulated data point corresponding to one source in our sample of 32 sources. We made 10,000 such groups, exhausting all the simulated data points. For each of these groups, we calculated Spearman correlation coefficients. The mean values of the correlation coefficients were taken as the representative values of our simulated distributions and the corresponding standard deviations as the uncertainties. Then the correlation coefficients were estimated to be $0.11 \pm 0.14$ for the overall gradient, $0.14 \pm 0.14$ for the rotational gradient, and $0.22 \pm 0.14$ for the normalized gradient. The fractions of the groups showing positive correlation coefficients are 0.77 for the overall gradient, 0.85 for the rotational gradient, and 0.94 for the normalized rotational gradient.

To test the significance of these correlations, we generated artificial uncorrelated samples. 
We randomly permutated our simulated velocity gradients with respect to our misalignment angles, and we repeated the same process as discussed above to obtain the distributions of the correlation coefficients for this random uncorrelated artificial data. We found that for the random data, the possibilities to obtain correlation coefficients greater than or equal to the mean correlation coefficients of our observed sample are $0.27$ for the overall gradient, $0.22$ for the rotational gradient, and $0.12$ for the normalized gradient. 
These values suggest that the normalized rotational gradient is most strongly correlated with the misalignment, followed by the rotational gradient and then the overall gradient.
This is in agreement with the trends observed in Figure \ref{fig:corr_deproj} and results from the original analysis (Section \ref{sec:InAnal}).
Our results suggest that the normalized rotational gradient is possibly correlated with the misalignment with a Spearman correlation of 0.22 and a confidence level of 88\% after considering the projection effect. Nevertheless, a larger sample is needed to have a more robust constraint on the correlation coefficient. 

\begin{figure}[htbp]
\centering
  \includegraphics[scale=0.8]{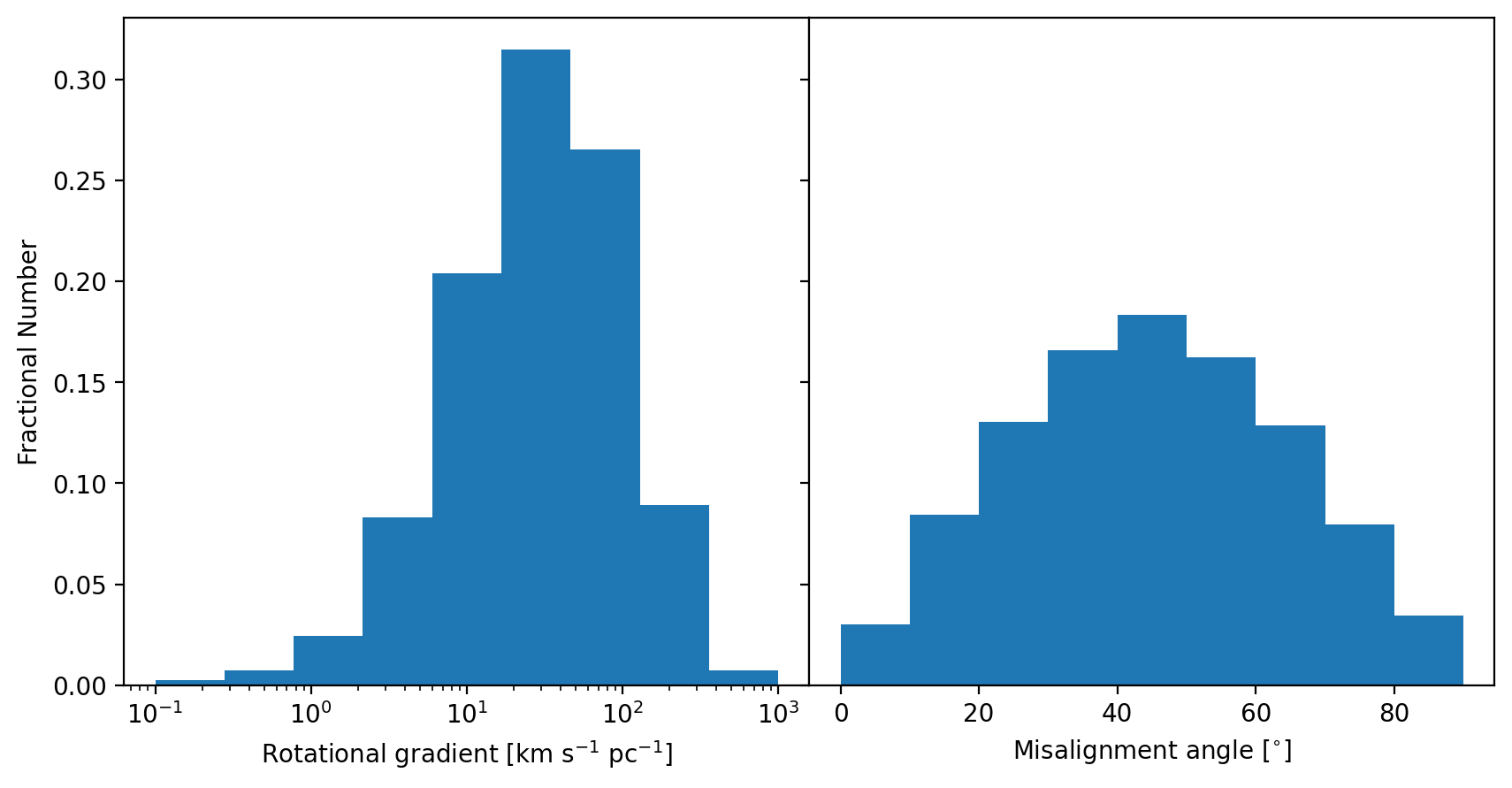}
    \caption{Histograms of the simulated deprojected rotational gradients in the protostellar envelopes (left panel, in a logarithmic scale) and deprojected misalignment angles between the core-scale magnetic fields and the outflow axes (right panel). These simulated probability distributions account for the measurement and projection uncertainties, as discussed in Section \ref{sec:FiAnal}.}
  \label{fig:dist}
\end{figure}

Moreover, we can see a characteristic range in the simulated probability distributions (innermost contours) for all the quantities. The distribution of the rotational gradients and misalignment angles, as shown in Figure \ref{fig:dist}, peaks around $\sim30$~km~s$^{-1}$~pc$^{-1}$ and $45 \arcdeg$, respectively. The distributions shown in Figure \ref{fig:dist} also include extra 22 sources with the velocity gradient measurements but without the magnetic field information and 7 more sources with the magnetic field information but without the velocity gradient measurements.
Not including these additional sources does not significantly change the final results. The characteristic range of the rotational gradients could be due to the underlying probability distribution of the angular momentum in the protostellar envelopes ($\sim1,000$~au), as discussed further in Section \ref{sec: VelDist}. 

Our polarization data are a subset of a larger sample of 62 sources from \citet{yen20}. \citet{yen20} identified dense cores in the JCMT $850~\micron$ maps using the clump identification algorithm \textit{clumpfind} \citep{wil94} and calculated misalignment angles within these detected cores. For different assumed distributions of misalignment angles in 3D space, they simulated distributions of projected angles and compared them with the observed distribution. Despite this slightly different characterization of core-scale magnetic field and handling of projection effects, \citet{yen20} inferred a similar distribution of deprojected angles, with more sources having intermediate misalignment angles ($\sim 30\arcdeg \textup{--} 60\arcdeg$). Also, our distribution of the deprojected angles depends on the assumed distributions of $\alpha$ and $\beta$ (Section \ref{sec:corrs}). Assuming other distributions of $\alpha$ and $\beta$ tends to shift the peak of the distribution of the misalignment angles towards larger values, 
as quantified in Appendix \ref{sec:alphabeta}.

\section{Discussions} \label{sec:discussion}

\subsection{Misalignment and angular momentum transportation} \label{sec:MisVSVel}

Ideal MHD simulations of collapse of dense cores with their rotational axes aligned with the magnetic fields, suggest that realistic levels of the magnetic field can greatly suppress rotation in the inner protostellar envelopes \citep{all03,gal06,mel08}. Further studies incorporated misalignment between the magnetic field and rotational axis of a dense core, and found that the misalignment can reduce the efficiency of magnetic braking and thus, allow inward angular momentum transportation to be more efficient in a collapsing dense core \citep{hen09,joo12,li13}. If the magnetic field orientation indeed significantly affects gas kinematics as suggested by these simulations, we expect to see stronger velocity gradients, particularly rotational gradients, for systems with larger misalignments.

However, as discussed in Section \ref{sec:InAnal}, we do not find any significant correlation between the overall or rotational gradients with respect to the misalignment. These results do not change even after accounting for the projection effects and the measurement uncertainties (Section \ref{sec:FiAnal}). This suggests that misalignment between the magnetic field and rotational axis in a dense core is not a dominant factor driving the gas kinematics or the amount of angular momentum at the envelope scale. 

This is different from the results of \citet{gal20}. For a sample of $\sim$20 protostars, \citet{gal20} measured magnetic field orientations in the protostellar envelopes on a scale of a few thousand au with SMA. They compared the misalignments between the magnetic fields and the outflow axes with the magnitudes of the overall velocity gradients at a $\sim5,000$~au scale, where some were taken from the literature \citep{wis01,sai99,tob11,tan11,gau20} and others were derived by \citet{gal20} using the published data \citep{mat08,hua13,tob18}. 
They primarily used the N$_{2}$H$^+$ emission to trace the gas kinematics.
They found a positive correlation between them with a Pearson correlation coefficient of 0.68, which could suggest greater misalignment results in greater angular momentum in protostellar envelopes. 

A subset of nine sources from \cite{gal20} are also in our sample. For these sources, the misalignment angles in \cite{gal20} are generally consistent with those derived from our data within the error bars, with a median difference of $11\arcdeg$. However, we found that our measured rotational gradients at a $1,000$~au scale are only weakly correlated with the velocity gradients in \citet{gal20}, which are at a larger scale, with a Spearman correlation coefficient of 0.41 and a corresponding p-value of 0.24.
Our velocity gradient measurements are typically greater by a factor of $\sim5$. 
 In addition, \citet{hem21} also used the same C$^{18}$O data from the MASSES survey to study the gas kinematics in the protostellar envelopes in our sample. They measured overall velocity gradients on variable scales ($\sim$1,000--3,000~au), depending on the sizes of the envelopes. We compared our overall velocity gradients to their measurements and found a strong correlation with a Spearman correlation coefficient of 0.59 and a p-value of 0.0003. Our velocity gradients measured at a 1,000 au scale are typically greater than their velocity gradients measured at a larger scale.
Thus, the discrepancy between the correlation observed by \citet{gal20} and no similar correlation found in our study could be due to the different spatial scales and underlying gas motions of the measured velocity gradients. Other observations have also found that the magnitudes and directions of velocity gradients in protostellar envelopes could change from large to small scales \citep{gau20}. In order to investigate this further, the correlations should be tested with a larger sample of sources with velocity gradient measurements at multiple scales.


Another factor influencing the amount of the angular momentum in a protostellar envelope can be the mass already accreted in its protostar-disk system. In the classical picture of collapse of a dense core, 
the internal distribution of the specific angular momentum is a increasing function of radius, and consequently,
the disk size increases with the enclosed mass of the central protostar-disk system \citep{ter84,bas98}. Similar trends have also been seen in non-ideal MHD simulations \citep{hen16,zha16,zha18}. 

The infalling motion in a protostellar envelope around a more massive protostar-disk system is expected to be faster because of its deeper gravitational potential, 
which could induce a larger velocity gradient along the outflow axis in the protostellar envelope \citep{yen13}.
Therefore, in order to delineate the role of the enclosed mass, we also normalized our rotational gradient by the velocity gradient along the outflow axis, and we found a significant correlation ($p\sim0.05$) between the normalized rotational gradient and the misalignment.
This observed correlation could suggest that for similar enclosed masses, more angular momentum is transported to protostellar envelopes in systems with greater misalignment. In other words, misalignment indeed could promote the amount of angular momentum transported to protostellar envelopes. However, it is not a dominant factor, and other parameters, like mass accretion in protostellar sources, also play an important role.

This is also in agreement with the non-ideal MHD simulations by \cite{hir20}. Along with misalignment angles, \cite{hir20} also varied the ratio of thermal-to-gravitational energy, taking it as a proxy for gravitational instability of the initial core. They found that the systems with smaller ratios form larger disks because the dense cores collapse more rapidly and gas is quickly advected to the disks. Moreover, they found that for systems with similar thermal-to-gravitational energy ratios, more misaligned magnetic field is conducive to form larger disks. 

    
\subsection{Distribution of Velocity Gradients} \label{sec: VelDist}

Figure \ref{fig:dist} (left panel) shows the distribution of the simulated deprojected rotational gradients for all the 54 sources. The median rotational gradient is $\sim 29$~km~s$^{-1}$~pc$^{-1}$ which at a radius of $1,000$~au corresponds to a specific angular momentum of $\sim 6.8\times 10^{-4}$~km~s$^{-1}$~pc. This value is in agreement with the mean specific angular momentum of $\sim 6\times 10^{-4}$~km~s$^{-1}$~pc at $<1600$~au scales for a sample of 12 protostars, inferred by \citet{gau20}.
Specific angular momenta of 17 Class 0 and I sources at $<1500$~au scales estimated by \citet{yen15} are also of the order of $\sim 10^{-4}$~km~s$^{-1}$~pc.

Assuming a given angular momentum is efficiently transported from this 1,000 au scale to the edge of a Keplerian disk with negligible mass compared to the stellar mass in an axisymmetric system without magnetic braking, the resultant disk radius can be estimated as \citep{ulr76,ter84,bas98}, 
\begin{equation}
    \label{equ:disk}
    R_{d} = \frac{l^{2}}{GM_{*}},
\end{equation}
where $R_{d}$ is the radius of the Keplerian disk, $l$ is the specific angular momentum, $G$ is the gravitational constant, and $M_{*}$ is mass of the central protostar. 

\citet{yen17} inferred a time-dependent mass accretion rate using bolometric luminosities and protostellar masses of a sample of 18 Class 0 and I protostars. We integrated this mass accretion rate over the typical lifetime for Class 0 sources of $0.26$~Myr \citep{dun15}. This gives a typical protostellar mass of $0.25$~M$_\odot$. Nevertheless, the masses of Class 0 protostars can still be different by two order of magnitude \citep{yen17}. 
Assuming the mass accretion rate is proportional to the protostellar mass, we can assume the distribution of masses of these young protostars to be similar to the mass distribution of main-sequence stars, i.e., the initial mass function (IMF). For low mass stars ($M_{*}\lesssim 1$~M$_\odot$), the IMF can be approximated as a log-normal distribution with a characteristic mass of $0.22$~M$_\odot$ and a variance of $\sim 0.57$~M$_\odot$ \citep{cha03}. As our sample Class 0 and I protostars continue to acquire more mass, we normalized this log-normal distribution of IMF to have a mean mass of $0.25$~M$_\odot$, instead of the original mean mass of $\sim 0.6$~M$_\odot$, and adopted this normalized distribution as the mass distribution of the young protostars.

By adopting the distributions of the deprojected rotational gradients and protostellar masses, we inferred the expected distribution of disk radii with Equation \ref{equ:disk}. We found the median disk radius to be $\sim 107$~au, comparable to the geometric mean ($10^{\mu(\log{R_d})}$, where $\mu$ is the simple mean) of $\sim 92$~au. The logarithmic variance ($\sigma(\log{R_d})$, where $\sigma$ is the simple variance) of the distribution is $\sim 1.4$.


\citet{tob20} observed the 0.87~mm continuum emission around 328 protostars in Orion clouds at a resolution of $\sim 40$~au. They reported a median dust disk radius of $\sim 48$~au for Class 0 sources and $\sim 38$~au for Class I sources. Among these only one source, B5-IRS1 has a disk radius $\gtrsim 100$~au. Similarly, \citet{enc21} surveyed the 0.87~mm continuum emission around 31 protostars in Ophiuchus at a resolution of $\sim 21$~au. They found the mean disk radius to be $\sim 24$~au for Class I sources and $\sim 17$~au for flat-spectrum sources. In Perseus, \citet{seg18} observed the $8$~mm continuum emission around 82 class 0 and I sources. With a resolution of $\sim 12$~au, they identified disk-like structures only around $22\%$ of the sources. However, \citet{seg18} also pointed out that the $8$~mm continuum emission traces large dust grains that could radially drift inwards and thus, these disk sizes are likely lower limits. As also shown in numerical simulations \citep{aso20}, continuum observations may not reliably trace entire Keplerian disks and can underestimate the disk size by a factor of $\sim2$--$3$. \citet{aso20} suggested that this is because as a disk grows, the density and temperature in the outer disk drops and the continuum emission becomes much fainter. 

For a small sample of young protostars, disks with radii of a few tens of au have been observed in molecular lines \citep{hsi19,tob12,tob20disk,rey21}.
\citet{mar20} used CO line observations at angular resolutions of $\sim0\farcs7$ to search for disks towards 16 nearby ($<500$~pc) Class 0 sources. They found clear Keplerian disks with radii $>50$~au in only two sources: L1448-C with a disk radius of $200$~au and L1527 with a disk radius of $90$~au.
Although the molecular-line measurements of Class 0 and I disks are still scarce, the results from \citet{mar20} suggest that only $\sim 13\%$ of disks have radii $\gtrsim100$~au. This ratio is much smaller compared to the disk radius distribution derived using our measurements of the angular momentum in the protostellar envelopes at a $1,000$~au scale, where $\sim50\%$ of disks are expected to be larger than $100$~au. 

One key assumption in deriving our expected distribution of disk radii is a conserved angular momentum within $\lesssim 1,000$~au scales.
Therefore, the apparent discrepancy between the derived and observed disk radii distributions is likely because the angular momentum is lost at $< 1,000$~au scales. This could be due to efficient magnetic braking in inner protostellar envelopes as a result of pinched field lines in this region \citep[e.g.,][]{li13}. A similar scenario has been observed in the Class I system HH 111, where the angular momentum in the envelope was observed to drop by a factor of $\sim3$ from 2000 au to 100 au scales \citep{chi16}. However, we note that protostellar envelopes with a relatively conserved angular momentum have also been observed \citep{aso15,aso17}.

Using measurements of the gas kinematics at $\sim1600$--$100$~au scales for 11 Class 0 protostars, \citet{gau20} identified eight sources with relatively flat (conserved) angular momentum profiles within this radial range.
Assuming that the angular momentum they measured at a $100$~au scale remains conserved till the edge of disks, they estimated expected disk radii for these source (using Equation \ref{equ:disk}). They found the estimated radii to be in agreement with the disk radii in those sources estimated with the continuum emission. However, it is important to note that \citet{gau20} compared the disk radii with the angular momentum at a $100$~au scale, different from our study with the angular momentum estimated at a $1,000$~au scale. Though, for these eight sources the angular momentum profiles were flatter than the other sources in their sample, still for most of their sample sources the angular momentum was observed to decrease from $1,000$ to $100$ au scales. Together with our results, the observations could suggest that the angular momentum is likely lost in protostellar envelops at radii between $\sim1,000$--$100$~au, possibly due to magnetic braking. In order to better characterize the scale-dependency of magnetic braking, a larger sample of young sources with resolved velocity profiles from envelopes to disks is needed.

\section{Conclusions} \label{sec:conclusion}

For a sample of 32 Class 0 and I protostars, we diagnosed the gas kinematics in the protostellar envelopes at a $1,000$~au scale using the C$^{18}$O data from the SMA MASSES Survey \citep{ste19} and the magnetic fields at the core scale of 4,000~au using the 850~$\micron$ polarimetric data from the JCMT BISTRO survey \citep{war17} and archive.
We assessed the overall, rotational, and infalling motions in the protostellar envelopes with the 2D velocity gradients, velocity gradients perpendicular to the outflows, and velocity gradients along the outflow axes in the C$^{18}$O moment 1 maps, respectively. 
We studied the dependence of the gas kinematics in the protostellar envelopes on the magnetic field structures in the dense cores, namely angular dispersion of the magnetic field and misalignment between the magnetic field and outflow axis (taken as a proxy for the rotational axis). Furthermore, we inferred an expected distribution of disk radii using the observed distribution of the rotational velocity gradients in the protostellar envelopes.
Our main results are:

\begin{enumerate}
    \item We did not find any significant correlation between the angles between the magnetic field and outflow axis in the dense cores at a 4,000~au scale, and the overall or rotational velocity gradients in the protostellar envelopes at a $1,000$~au scale. We also did not find any correlation between the angular dispersions of the magnetic fields and the velocity gradients. These results could suggest that the misalignment between the magnetic field and rotational axis and the ratio of the turbulence to the magnetic field strength in a dense core are not dominant factors in determining gas kinematics or angular momentum in its protostellar envelope.  
    \item We found a significant correlation between the rotational velocity gradients normalized by the infalling velocity gradients in the protostellar envelope and the misalignment angles between the magnetic fields and outflows in the dense cores. 
    In particular, these normalized values transition from smaller-than one to larger-than one, suggesting the presence of an infall-dominated regime with small misalignment angles and a rotation-dominated regime with larger misalignment.
    The Spearman correlation coefficient was calculated to be 0.35 with a p-value of 0.05. After considering projection effects, the Spearman correlation coefficient becomes 0.22 with a confidence level of 88\%, which are related to the assumed probability distributions of 3D orientations of the magnetic field and outflows. Assuming that the infalling velocity is proportional to the mass of a central protostar-disk system, our results could suggest that for similar central masses, more angular momentum is transported to protostellar envelopes in systems with greater misalignment. This hints that misalignment between the magnetic field and rotational axis in a dense core could promote angular momentum transportation from large to small scales, although it is not a dominant factor. 
    \item Assuming our estimated angular momentum in the protostellar envelopes at a $1,000$~au scale is efficiently transported to disk-forming regions, the median disk radius is expected to be $\sim100$~au. However, molecular-line observations like in \citet{mar20} show that disks with radii $\gtrsim100$~au are not common for Class 0 and I sources. Thus, this suggests that the angular momentum is likely lost in protostellar envelops at radii between $\sim1,000$--$100$~au, possibly due to magnetic braking.
\end{enumerate}

\begin{acknowledgments}
The Submillimeter Array is a joint project between the Smithsonian Astrophysical Observatory and the Academia Sinica Institute of Astronomy and Astrophysics and is funded by the Smithsonian Institution and the Academia Sinica.
The James Clerk Maxwell Telescope is operated by the East Asian Observatory on behalf of The National Astronomical Observatory of Japan; Academia Sinica Institute of Astronomy and Astrophysics; the Korea Astronomy and Space Science Institute; the National Astronomical Research Institute of Thailand; Center for Astronomical Mega-Science (as well as the National Key R\&D Program of China with No.~2017YFA0402700). Additional funding support is provided by the Science and Technology Facilities Council of the United Kingdom and participating universities and organizations in the United Kingdom and Canada. 
H.-W.Y.\ acknowledges support from the Ministry of Science and Technology (MOST) in Taiwan through grant MOST 108-2112-M-001-003-MY2 and MOST 110-2628-M-001-003-MY3 and support from an Academia Sinica Career Development Award.
P.M.K.\ is supported by the Ministry of Science and Technology (MoST) through grants MoST 109-2112-M-001-022 and MoST 110-2112-M-001-057.  
E.J.C.\ was supported by the National Research Foundation of Korea (NRF) grant funded by the Korea government (MSIT) (No.~NRF-2019R1I1A1A01042480).
J.K.\ is supported by JSPS KAKENHI grant No.~19K14775.
W.K.\ was supported by the National Research Foundation of Korea (NRF) grant funded by the Korea government (MSIT) (NRF-2021R1F1A1061794).
C.W.L.\ is supported by the Basic Science Research Program through the National Research Foundation of Korea (NRF) funded by the Ministry of Education, Science and Technology (NRF-2019R1A2C1010851).
K.P.\ is a Royal Society University Research Fellow, supported by grant number URF\textbackslash R1\textbackslash211322.
M.T.\ is supported by JSPS KAKENHI grant Nos.~18H05442, 15H02063,and 22000005. 

\end{acknowledgments}

\appendix

\section{Position-Velocity Diagrams} \label{sec:PVs}
\begin{figure}[htbp]
\centering
  \includegraphics[scale=0.20]{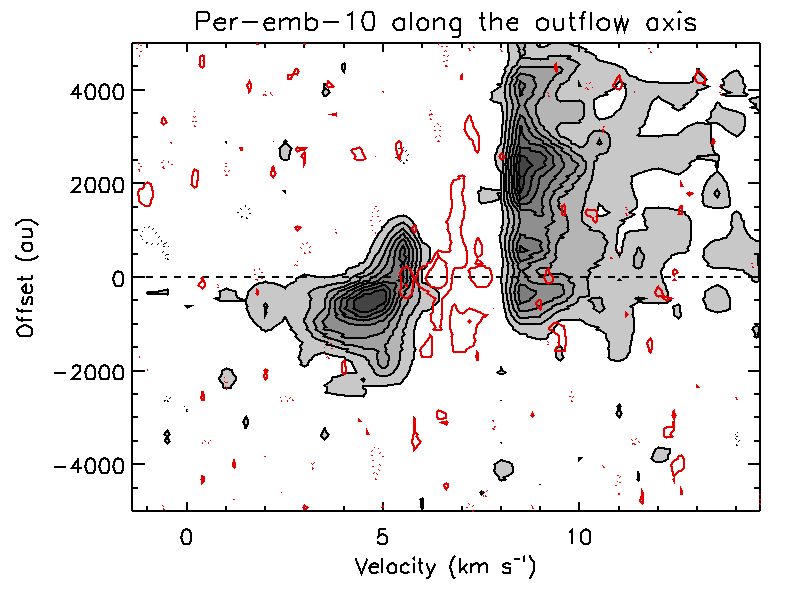}
  \includegraphics[scale=0.20]{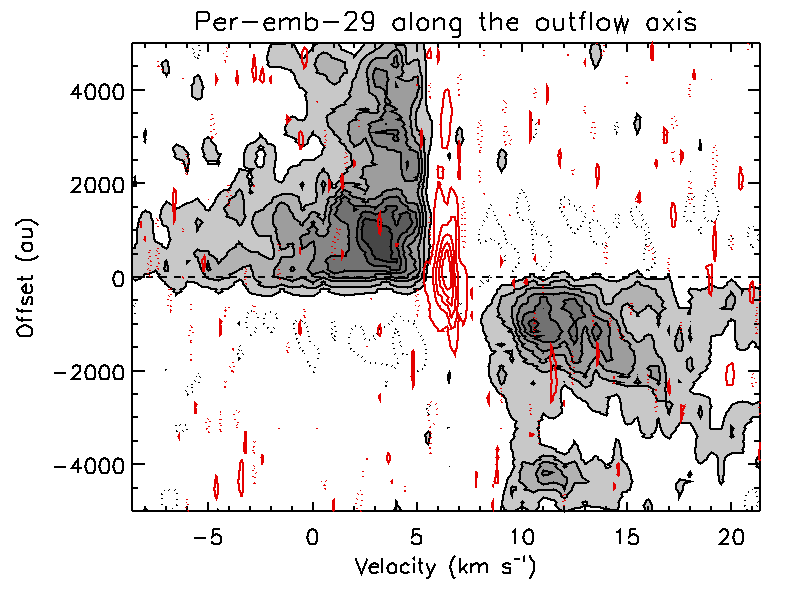}
  \includegraphics[scale=0.20]{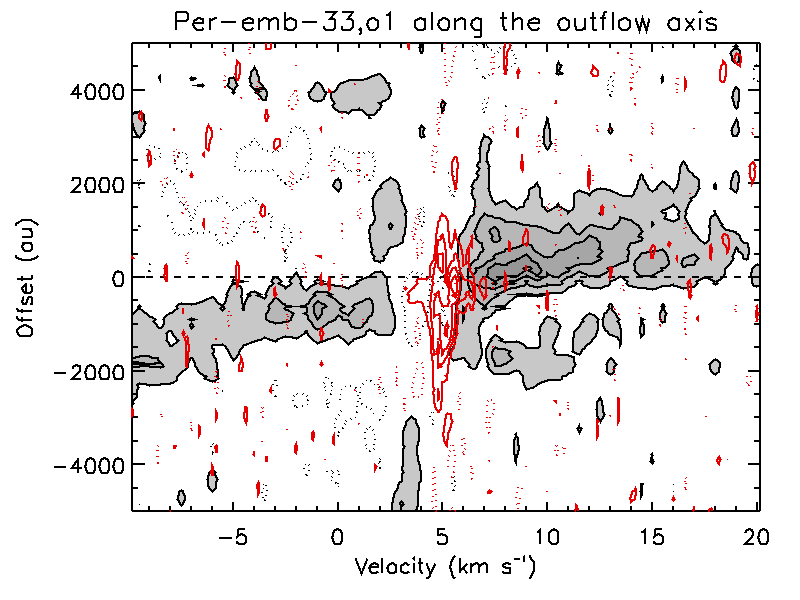}
    \caption{ Position-Velocity Diagrams of the C$^{18}$O ($2-1$) emission (red contours) and CO ($3-2$) emission (grey contours), along the outflow axes in Per-emb-10 (left), Per-emb-29 (middle), and Per-emb-33~O1 (right). Contour levels for the C$^{18}$O emission are from 2$\sigma$ in steps of 2$\sigma$, 5$\sigma$, and 3$\sigma$ for Per-emb-10, Per-emb-29, and Per-emb-33~O1, respectively. Contour levels for the CO emission are from 3$\sigma$ in steps of 3$\sigma$, 3$\sigma$, and 5$\sigma$ for Per-emb-10, Per-emb-29, and Per-emb-33~O1, respectively. All the PV diagrams are centered at the protostellar positions.}
  \label{fig:pv2}
\end{figure}

Figure \ref{fig:pv2} compares the PV diagrams along the outflow axis of the C$^{18}$O ($2-1$) and CO ($3-2$) emission in Per-emb-10, Per-emb-29, and Per-emb-33~O1, as examples. The CO ($3-2$) emission primarily traces the molecular outflows \citep{ste19} and shows comparable line widths or increasing velocities from small to large radii. Different from the CO ($3-2$) emission, in the PV diagrams along the outflow axes, the C$^{18}$O ($2-1$) emission at higher velocities tends to appear close to the protostellar positions.

\begin{figure}[htbp]
\centering
  \includegraphics[scale=0.26]{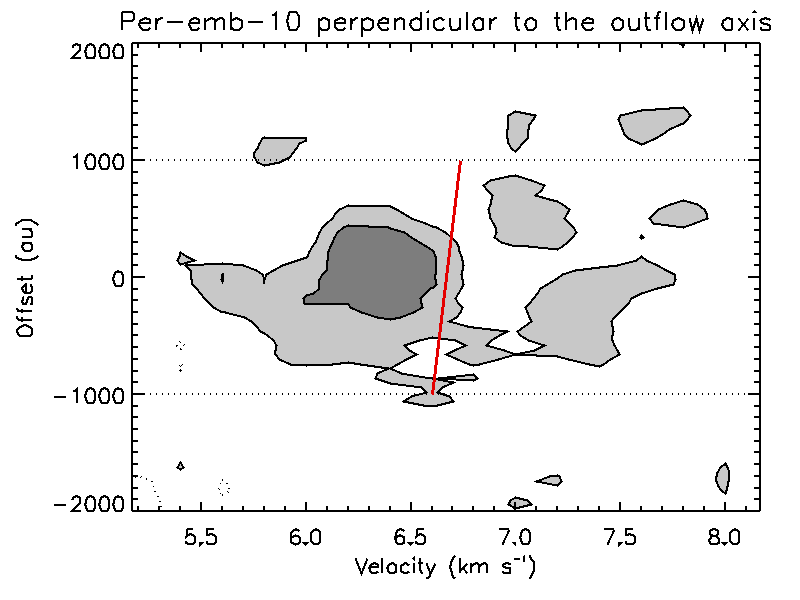}
  \includegraphics[scale=0.26]{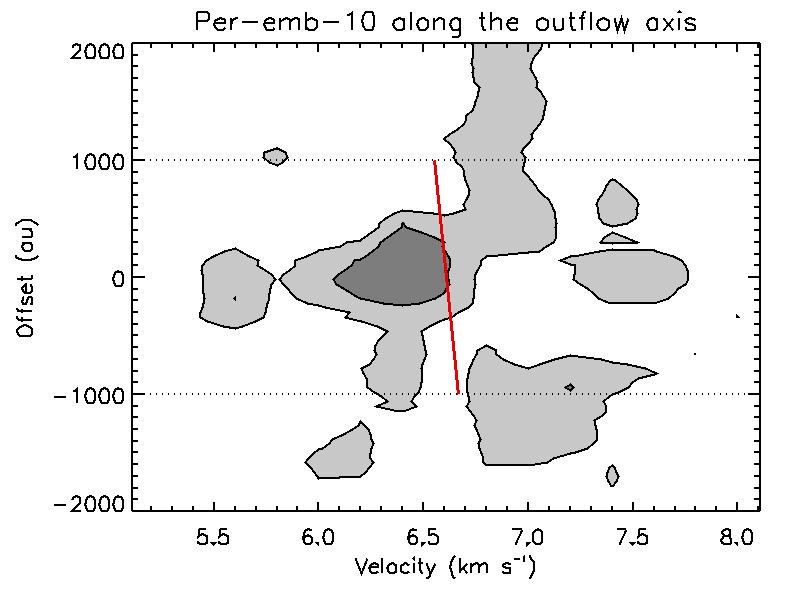}
  \includegraphics[scale=0.26]{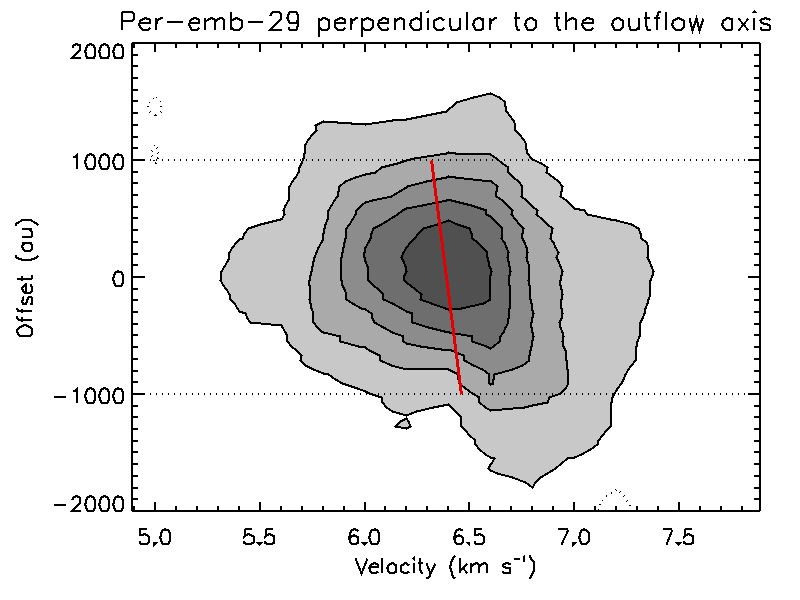}
  \includegraphics[scale=0.26]{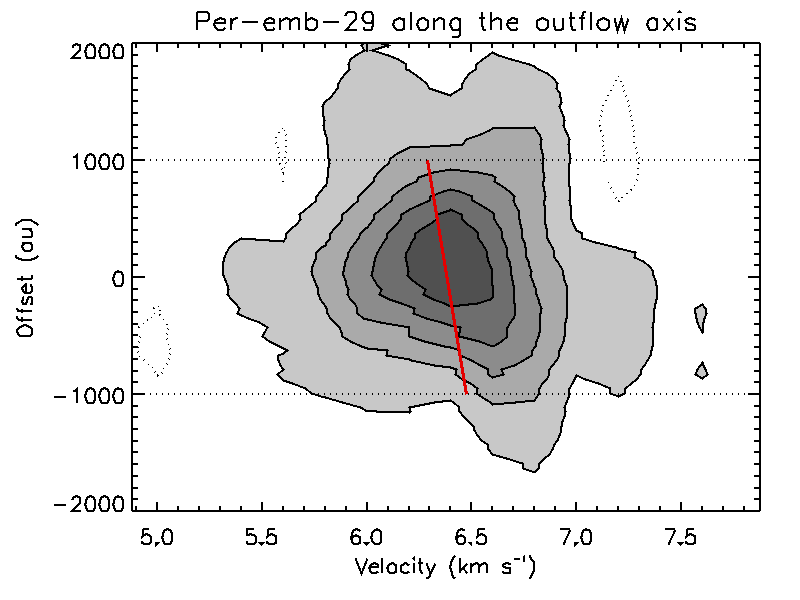}
  \includegraphics[scale=0.26]{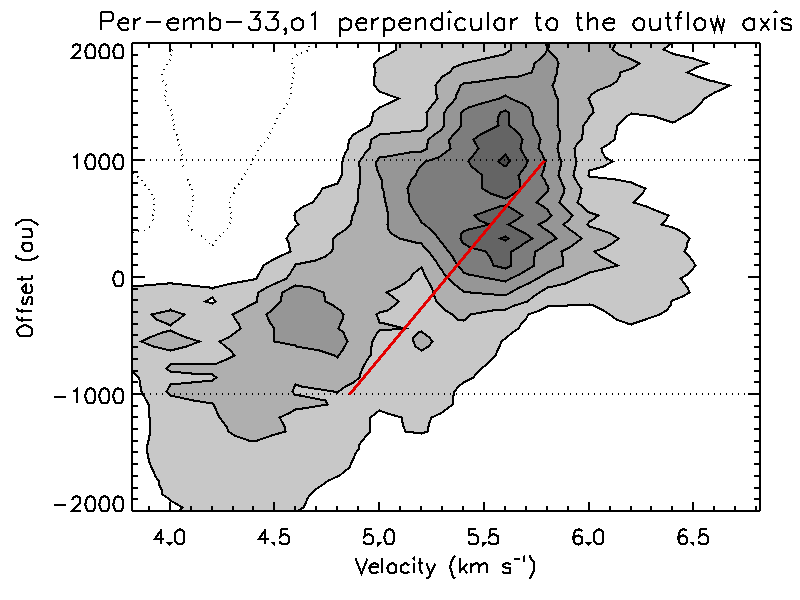}
  \includegraphics[scale=0.26]{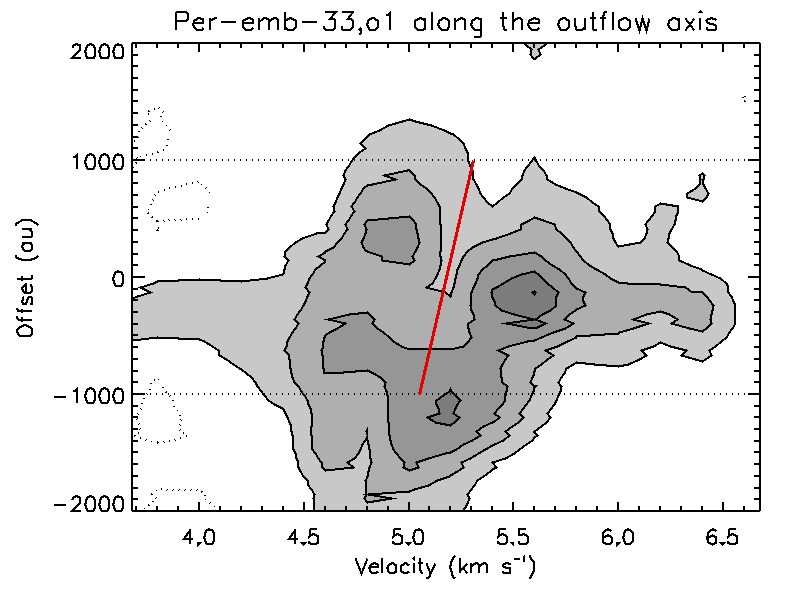}
    \caption{Position-Velocity Diagrams of the C$^{18}$O emission perpendicular (left column) and along (right column)} the outflow axes in Per-emb-10 (top row), Per-emb-29 (middle row), and Per-emb-33~O1 (bottom row). Contour levels are from 2$\sigma$ in steps of 2$\sigma$, 5$\sigma$, and 3$\sigma$ in the top, middle, and bottom rows, respectively. Red solid lines delineate the velocity gradients along and perpendicular to the outflow axes measured from the moment 1 maps of the C$^{18}$O emission in these sources. Dotted horizontal lines enclose the central region within a radius of 1,000 au adopted to measure the velocity gradients. All the PV diagrams are centered at the protostellar positions.
  \label{fig:pv}
\end{figure}

Figure \ref{fig:pv} presents the PV diagrams of the C$^{18}$O emission along and perpendicular to the outflow axes, along with the velocity gradients measured from the moment 1 maps (Section \ref{sec:kinmatics}). The observed velocity structures in the PV diagrams indeed can be approximately described with the measured velocity gradients.

\section{Velocity Gradients and Magnetic Field Measurements} \label{sec:measurements}

Table \ref{tab:measurements} presents all the measurements used in this paper.

\begin{longrotatetable} 
\movetabledown=25mm
\begin{deluxetable}{ccccccccccccccccc} \label{tab:measurements}
    \tabletypesize{\scriptsize}
    \tablecaption{Velocity Gradients and Magnetic Field Measurements}
    \tablehead{\colhead{Source} & \colhead{R.A.\,(J2000)} & \colhead{Decl.\,(J2000)} & \colhead{Outflow P.A.} & \colhead{$VG_{\rm 2D}$} & \colhead{$\Delta VG_{\rm 2D}$} & \colhead{$\theta_{\rm VG}$} & \colhead{$\Delta \theta_{\rm VG}$} & \colhead{$VG_{\rm rot}$} & \colhead{$\Delta VG_{\rm rot}$} & \colhead{$VG_{\rm inf}$} & \colhead{$\Delta VG_{\rm inf}$} & \colhead{$\theta_{\rm B}$} & \colhead{$\Delta \theta_{\rm B}$} & \colhead{Misalignment} & \colhead{$\delta \theta_{\rm B}$} & \colhead{$\Delta \delta \theta_{\rm B}$}\\ 
    \colhead{} & \colhead{$\mathrm{{}^{\circ}}$} & \colhead{$\mathrm{{}^{\circ}}$} & \colhead{$\mathrm{{}^{\circ}}$} & \colhead{$\mathrm{km\,pc^{-1}\,s^{-1}}$} & \colhead{$\mathrm{km\,pc^{-1}\,s^{-1}}$} & \colhead{$\mathrm{{}^{\circ}}$} & \colhead{$\mathrm{{}^{\circ}}$} & \colhead{$\mathrm{km\,pc^{-1}\,s^{-1}}$} & \colhead{$\mathrm{km\,pc^{-1}\,s^{-1}}$} & \colhead{$\mathrm{km\,pc^{-1}\,s^{-1}}$} & \colhead{$\mathrm{km\,pc^{-1}\,s^{-1}}$} & \colhead{$\mathrm{{}^{\circ}}$} & \colhead{$\mathrm{{}^{\circ}}$} & \colhead{$\mathrm{{}^{\circ}}$} & \colhead{$\mathrm{{}^{\circ}}$} & \colhead{$\mathrm{{}^{\circ}}$}}
    \startdata
    Per-emb-1 & 55.98669 & 32.01394 & 116 & 20.9 & 1.5 & -109.9 & 4.1 & 3.9 & 3.0 & 19.5 & 3.4 & 163.2 & 0.8 & 47.2 & 7.2 & 0.8 \\
    Per-emb-3 & 52.2524 & 31.20006 & 97 & 22.0 & 5.2 & -105.6 & 17.7 & 15.6 & 28.6 & 44.6 & 11.4 & 43.9 & 7.4 & 53.1 & 22.9 & 5.4 \\
    Per-emb-6 & 53.31002 & 31.11964 & 60 & 22.0 & 4.8 & -11.1 & 12.1 & 31.0 & 8.5 & 38.9 & 14.0 & 149.7 & 3.0 & 89.7 & 6.5 & 2.8 \\
    Per-emb-10 & 53.31843 & 31.11446 & 52 & 36.5 & 8.0 & -5.9 & 15.1 & 13.9 & 14.8 & 11.9 & 26.2 & 163.1 & 4.3 & 68.9 & 4.5 & 4.4 \\
    Per-emb-11,O1 & 55.98777 & 32.05133 & 162 & 19.1 & 8.6 & -151.7 & 33.8 & 25.9 & 8.4 & 130.2 & 7.4 & 155.1 & 3.7 & 6.9 & 3.8 & 5.0 \\
    Per-emb-11,O2 & 55.99037 & 32.05277 & 36 & 28.3 & 6.0 & 92.6 & 12.2 & 26.8 & 6.2 & 26.8 & 9.1 & 155.3 & 2.1 & 60.7 & 1.5 & 2.0 \\
    Per-emb-12 & 52.2939 & 31.22526 & 35 & 15.6 & 3.5 & 54.2 & 13.0 & 62.5 & 1.7 & 7.4 & 6.9 & 59.3 & 0.1 & 24.3 & 1.5 & 0.1 \\
    Per-emb-13,O1 & 52.30007 & 31.2189 & 180 & 25.7 & 2.4 & 103.2 & 5.2 & 30.0 & 2.4 & 2.7 & 2.3 & 70.1 & 0.9 & 70.1 & 3.6 & 0.9 \\
    Per-emb-13,O2 & 52.30351 & 31.21858 & 90 & 29.6 & 8.0 & -89.2 & 10.4 & 9.6 & 11.4 & 18.7 & 22.8 & 82.0 & 1.5 & 8.0 & 7.9 & 1.5 \\
    Per-emb-15 & 52.2669 & 31.24618 & 145 & 23.6 & 0.3 & -19.8 & 0.8 & 18.8 & 1.2 & 14.0 & 0.4 & 179.6 & 3.0 & 34.6 & 12.8 & 3.3 \\
    Per-emb-16 & 55.96241 & 32.05669 & 11 & 35.5 & 4.2 & 153.9 & 7.6 & 44.2 & 12.2 & 69.5 & 8.7 & 105.2 & 5.2 & 85.8 & 11.9 & 5.5 \\
    Per-emb-17 & 51.91293 & 30.21752 & 57 & 31.2 & 6.9 & 25.6 & 12.6 & 143.0 & 21.3 & 1.9 & 17.7 & 71.8 & 5.5 & 14.8 & 11.2 & 5.8 \\
    Per-emb-18 & 52.29691 & 31.30863 & 150 & 6.8 & 3.6 & 132.5 & 29.9 & 36.4 & 9.5 & 13.6 & 5.1 & 101.3 & 4.1 & 48.7 &   &   \\
    Per-emb-20 & 51.93032 & 30.20799 & 115 & 21.1 & 1.1 & 101.3 & 3.0 & 10.8 & 2.7 & 21.8 & 3.4 & 96.5 & 3.5 & 18.5 & 1.9 & 3.4 \\
    Per-emb-21 & 52.29445 & 31.30561 & 48 & 20.4 & 10.4 & 177.9 & 29.2 & 77.4 & 23.8 & 17.7 & 14.0 & 96.7 & 1.4 & 48.7 & 4.8 & 1.5 \\
    Per-emb-22 & 51.34338 & 30.75368 & 118 & 30.1 & 4.3 & -77.9 & 8.2 & 6.4 & 3.2 & 44.7 & 5.7 & 123.8 & 2.6 & 5.8 & 9.3 & 2.1 \\
    Per-emb-27,O1 & 52.23154 & 31.24362 & 14 & 31.0 & 2.5 & -75.4 & 4.6 & 61.5 & 5.6 & 8.0 & 4.1 & 72.4 & 1.3 & 58.4 & 7.0 & 1.7 \\
    Per-emb-27,O2 & 52.23151 & 31.24345 & 104 & 23.3 & 2.7 & -86.6 & 6.7 & 2.1 & 3.7 & 54.6 & 3.7 & 71.4 & 1.2 & 32.6 & 6.8 & 1.5 \\
    Per-emb-28 & 55.96253 & 32.05223 & 112 & 25.4 & 7.3 & 178.6 & 17.0 & 24.8 & 8.1 & 52.2 & 7.2 & 118.5 & 6.3 & 6.5 & 25.6 & 5.1 \\
    Per-emb-29 & 53.32449 & 31.15884 & 132 & 25.5 & 1.4 & 103.9 & 3.1 & 14.7 & 3.8 & 19.1 & 3.8 & 100.5 & 1.5 & 31.5 & 5.9 & 2.1 \\
    Per-emb-33,O1 & 51.40158 & 30.75409 & 122 & 116.9 & 4.4 & -48.8 & 2.2 & 95.8 & 10.9 & 26.5 & 10.9 & 38.9 & 2.7 & 83.1 & 4.7 & 2.7 \\
    Per-emb-33,O2 & 51.40208 & 30.75608 & 38 & 68.2 & 4.3 & -60.2 & 3.7 & 83.7 & 11.5 & 60.8 & 3.3 & 33.4 & 4.7 & 4.6 & 31.4 & 3.2 \\
    Per-emb-33,O3 & 51.39862 & 30.75948 & 128 & 136.5 & 4.8 & -43.2 & 2.0 & 134.0 & 17.7 & 31.4 & 10.4 & 165.2 & 2.3 & 37.2 & 11.8 & 4.9 \\
    Per-emb-35,O1 & 52.15455 & 31.22522 & 123 & 18.7 & 0.9 & 151.7 & 2.6 & 15.1 & 1.7 & 4.7 & 2.8 & 89.2 & 4.7 & 33.8 & 12.7 & 4.3 \\
    Per-emb-35,O2 & 52.15508 & 31.22549 & 169 & 13.0 & 0.9 & 120.2 & 4.1 & 14.3 & 2.1 & 5.4 & 3.3 & 89.2 & 4.7 & 79.8 & 12.7 & 4.3 \\
    Per-emb-36 & 52.23906 & 31.23771 & 24 & 30.1 & 3.6 & -112.4 & 6.7 & 46.4 & 7.0 & 8.6 & 9.9 & 85.4 & 3.3 & 61.4 & 6.7 & 2.8 \\
    Per-emb-40 & 53.31945 & 31.13192 & 101 & 43.8 & 2.8 & -0.2 & 3.7 & 61.6 & 4.0 & 14.0 & 2.6 & 155.5 & 2.6 & 54.5 & 2.6 & 2.6 \\
    Per-emb-42 & 51.41306 & 30.73275 & 43 & 44.9 & 2.8 & 155.0 & 3.5 & 48.9 & 5.5 & 36.9 & 3.8 & 174.3 & 9.5 & 48.7 &   &   \\
    Per-emb-44 & 52.26569 & 31.26772 & 130 & 30.8 & 3.2 & -143.3 & 5.8 & 2.4 & 3.9 & 60.1 & 2.6 & 159.9 & 1.1 & 29.9 & 0.9 & 2.0 \\
    Per-emb-58 & 52.24342 & 31.37152 & 167 & 15.1 & 0.7 & 95.7 & 2.7 & 18.0 & 0.7 & 7.8 & 0.5 & 122.9 & 3.1 & 44.1 & 4.7 & 3.4 \\
    SVS 13B & 52.26282 & 31.26437 & 170 & 13.3 & 1.5 & -165.4 & 6.3 & 5.1 & 4.9 & 13.0 & 2.5 & 165.2 & 1.0 & 4.8 & 6.8 & 1.1 \\
    SVS 13C & 52.25821 & 31.26057 & 8 & 31.3 & 3.0 & 86.5 & 5.4 & 47.4 & 2.0 & 33.8 & 3.4 & 161.6 & 1.9 & 26.4 & 10.7 & 4.0
    \enddata
    \tablecomments{$VG_{\rm 2D}$,$VG_{\rm rot}$, and $VG_{\rm inf}$ are overall velocity gradients, velocity gradients perpendicular to the outflow axis, and velocity gradients parallel to the outflow axis, respectively. $\theta_{\rm VG}$ and $\theta_{\rm B}$ are the orientations of overall velocity gradients and 4,000-au scale magnetic fields from north to east. Misalignment refers to the misalignment angles between magnetic fields and outflows. $\delta \theta_{\rm B}$ are angular dispersions in magnetic fields at a 4,000~au scale. $\Delta$ denotes measurement uncertainties in the respective quantities. Distances were taken from \cite{zuc18}.}
\end{deluxetable}
\end{longrotatetable} 


\section{Correction for Projection effects of angles} \label{sec:angles}

\begin{figure}[htbp]
\centering
  \includegraphics[scale=0.8]{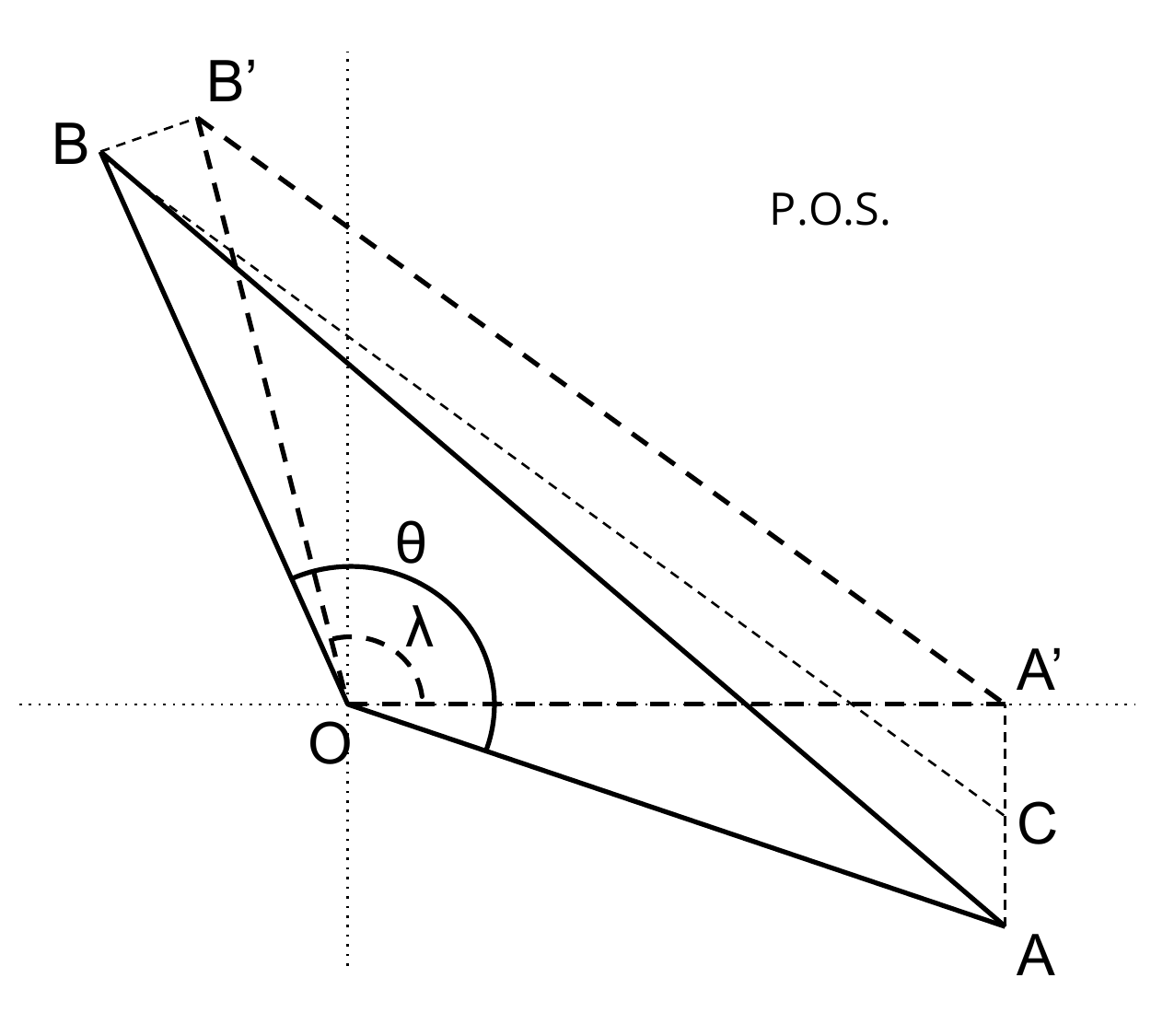}
    \caption{Schematic diagram of the projection effect. OA and OB represent actual vector orientations in 3D whereas OA' and OB' represent projections of these vectors on POS.}
  \label{fig:3d}
\end{figure}

Our measured misalignments between the outflows and magnetic fields in the dense cores are angles projected on  POS, and the actual misalignment angles in 3D can differ significantly, as discussed in Section \ref{sec:corrs}. In order to account for these projection effects, we created a geometric model of the vectors in 3D projected on a 2D surface, as shown in Figure \ref{fig:3d}. 

Assuming OA to be parallel to the rotational axis, OB to be parallel to the magnetic field, and OA$^\prime$ and OB$^\prime$ to be their projections on POS (i.e., $\angle BB^{\prime}A^\prime = \angle AA^{\prime}B^{\prime} = 90\arcdeg$). Then, the observed misalignment will be $\angle$A$^\prime$OB$^\prime$ or $\lambda$, and the actual misalignment will be $\angle$AOB or $\theta$. Here we define $\angle$AOA$^\prime$ and $\angle$BOB$^\prime$, i.e., inclinations of AO and BO with respect to POS, are $\alpha$ and $\beta$, respectively. For simplicity, we assume $|OA| = |OB| = r$. This is equivalent to working with unity vectors which is sufficient here because we are only concerned about deprojecting directions but not the magnitude of vectors. Using the trigonometric relations, we derive the following,
\begin{align*}
    |AA^\prime| = r \sin{\alpha}, \\
    |OA^\prime| = r \cos{\alpha}, \\
    |BB^\prime| = r \sin{\beta}, \\
    |OB^\prime| = r \cos{\beta}.
\end{align*}
Now in $\triangle AOB$,
\begin{align} \label{equ:len1}
\begin{split}
    |AB|^2 & = r^2 + r^2 - 2 r^2 \cos{\theta} \\
    &= 2 r^2 (1-\cos{\theta}). 
\end{split}
\end{align}
Similarly in $\triangle A^{\prime}OB^\prime$,
\begin{align} \label{equ:len2}
\begin{split}
    |A^{\prime}B^\prime|^2 & = (r \cos{\alpha})^2 + (r \cos{\beta})^2 - 2 (r \cos{\alpha}) (r \cos{\beta}) \cos{\lambda} \\
    & = r^2 (\cos^2{\alpha} + \cos^2\beta - 2 \cos{\alpha} \cos{\beta} \cos{\lambda}).
\end{split}
\end{align}

Let C be a point on AA$^\prime$ such that CB $||$ A$^\prime$B$^\prime$. Because $BB^\prime \bot B^{\prime}A^\prime$ and $CA^\prime \bot B^{\prime}A^\prime$, 
\begin{align*}
    |CB| &= |A^{\prime}B^\prime|, \\
    |AC| &= |AA^\prime| - |BB^\prime|, \\
    \angle ACB &= \angle CA^{\prime}B^\prime = 90\arcdeg.
\end{align*}
Applying the Pythagoras theorem in $\triangle ABC$:
\begin{align}
    |AB|^2 &= |AC|^2 + |CB|^2. \nonumber \\
    \text{From Equation \ref{equ:len1} and \ref{equ:len2},} \nonumber \\
    2 r^2 (1-\cos{\theta}) &= (r\sin{\alpha} - r \sin{\beta})^2 + r^2 (\cos^2{\alpha} + \cos^2\beta - 2 \cos{\alpha} \cos{\beta} \cos{\lambda}) \nonumber \\
    2-2\cos{\theta} &= \sin^2{\alpha} + \sin^2{\beta} - 2\sin{\alpha}\sin{\beta} + \cos^2{\alpha} + \cos^2\beta - 2 \cos{\alpha} \cos{\beta} \cos{\lambda} \nonumber \\
    \cos{\theta} &= \sin{\alpha}\sin{\beta} + \cos{\alpha} \cos{\beta} \cos{\lambda}. \label{equ:ProjEqu}
\end{align}
The Equation \ref{equ:ProjEqu} expresses the actual misalignment in 3D ($\theta$) as a function of the observed misalignment ($\lambda$), the inclination of the rotational axis with respect to POS ($\alpha$), and the inclination of the magnetic field with respect to POS ($\beta$).

\section{Different Distributions of inclination angles} \label{sec:alphabeta}


Table \ref{tab:AngDist} gives the median correlation coefficients and the median misalignment angles in 3D for different assumed probability distributions of the inclination angles, $\alpha$ and $\beta$. We note that the correlation coefficient between the normalized rotational gradient and the misalignment become almost zero when the uniform distributions of $\alpha$ and $\beta$ ranging from  $-90\arcdeg$ to $+90\arcdeg$ are adopted. Nevertheless, these distributions of $\alpha$ and $\beta$, which imply more magnetic fields and outflows are along the line of slight are less likely in our sample, and are against the criteria of our sample selection, which are based on the detections of magnetic fields and outflows projected on POS (as discussed in Section \ref{sec:corrs}).



\begin{deluxetable}{ccccc} \label{tab:AngDist}
    \tablecaption{Results for different assumed distributions of inclinations}
    \tablehead{\colhead{Distribution} & \colhead{R(VG$_{ove}$)} & 
    \colhead{R(VG$_{rot}$)} & 
    \colhead{R(VG$_{rot}$/VG$_{inf}$)} & 
    \colhead{Median 3D Misalignment}} 
    \startdata
    Cosine &  0.11 &  0.14 &  0.22 &  43$\arcdeg$\\
    Normal ($\mu=0\arcdeg$, $\sigma=30\arcdeg$) &  0.02 &  0.16 &  0.21 &  56$\arcdeg$ \\
    Uniform ($-60\arcdeg$--$+60\arcdeg$) &  0.03 &  0.11 &  0.15 &  58$\arcdeg$ \\
    Uniform ($-90\arcdeg$--$+90\arcdeg$) &  0.03 &  -0.05 &  0.07 &  55$\arcdeg$ \\ 
    \enddata
    \tablecomments{R(VG$_{ove}$), R(VG$_{rot}$), and R(VG$_{rot}$/VG$_{inf}$) are the median Spearman correlation coefficients for the overall, rotational, and normalized rotational velocity gradients, respectively.}
\end{deluxetable}

\bibliographystyle{aasjournal}
\bibliography{references}{}
\end{document}